\documentclass{article}
\usepackage{amssymb}

\newtheorem{theorem}{Theorem}
\newtheorem{proposition}{Proposition}
\newtheorem{lemma}{Lemma}
\newtheorem{remark}{Remark}

\begin{document}
\title{On the Law of Addition of Random Matrices}

\author{L. Pastur$^{1,3}$\thanks{\emph{On leave from the
U.F.R. de Math{\'e}matiques, Universit{\'e}
Paris 7} } \and V. Vasilchuk$^{2,3}$}

\date{}
\maketitle
\begin{center}\begin{small}
{$^1$ Centre de Physique Th{\'e}orique de CNRS, Luminy - case 907,}
{13288, Marseille, France, E-mail: pastur@cpt.univ-mrs.fr  }

{$^2$ U.F.R. de Math{\'e}matiques, Universit{\'e} Paris 7, 2, place Jussieu,}
{75251, Paris \\ Cedex 05, France}

{$^3$ Mathematical Division, Institute for Low Temperature Physics,}
{47, Lenin Ave., 310164, Kharkov, Ukraine, E-mail: vasilchuk@ilt.kharkov.ua}
\end{small}
\end{center}

\begin{abstract}
\noindent Normalized eigenvalue counting measure of the sum of two Hermitian
(or real symmetric) matrices $A_{n}$ and $B_{n}$ rotated independently with
respect to each other by the random unitary (or orthogonal) Haar distributed
matrix $U_{n}$ (i.e. $A_{n}+U_{n}^{\ast }B_{n}U_{n}$) is studied in the
limit of large matrix order $n$. Convergence in probability to a limiting
nonrandom measure is established. A functional equation for the Stieltjes
transform of the limiting measure in terms of limiting eigenvalue measures
of $A_{n}$ and $B_{n}$ is obtained and studied.

\end{abstract}

\section{Introduction}
The paper deals with the eigenvalue distribution of the sum of two $n\times
n $ Hermitian or real symmetric random matrices as $n\rightarrow \infty $.
Namely we express the limiting normalized counting measure of eigenvalues of
the sum via the same measures of its two terms, assuming that latter exist
and that terms are randomly rotated one with respect another by an unitary
or an orthogonal random matrix uniformly distributed over the group $U(n)$
or $O(n)$ respectively.

One may mention several motivations of the problem. First, it can be
regarded in the context of general problem to describe the eigenvalues of
the sum of two matrices in terms of eigenvalues of two terms of the sum. The
latter problem dates back at least to the paper of H. Weyl \cite{We:12}, was
treated in a number of papers, including the recent paper \cite{Kl:98}, and
related to interesting questions of combinatorics, geometry, algebra etc.
(see e.g. \cite{Fu:98} for recent results and references). The problem is
also of considerable interest for mathematical physics because of its
evident links with spectral theory and quantum mechanics (perturbation
theory in particular).

It is clear that one cannot expect in general a simple and closed expression
for eigenvalues of the sum of two given matrices via eigenvalues of terms.
Hence, it is natural to look for a ``generic'' asymptotic answer, studying a
randomized version of the problem in which at least one of the two terms is
random and both behave rather regularly as $n\rightarrow \infty $.
Particular results of this type were given in \cite{Ma-Pa:67,Pa:72} where it
was proved that under certain conditions the divided by $n$ eigenvalue
counting measure of the sum converges in probability to the nonrandom limit
that can be found as a unique solution of a certain functional equation.
Thus, a randomized version of the problem admits a rather constructive and
explicit solution. These results were developed in several directions (see
e.g. \cite{Gi:75} - \cite{KKP:96} and the recent work \cite{Pa:99}).
Similar problems arose recently in operator algebras studies, known now as
the free (non-commutative) probability (see \cite{Vo:91,Vo-Dy-Ni:92,Vo:97}
for results and references). In particular, the notion of the $R$%
-transform and the free convolution of measures were introduced by Voiculescu
and allowed the limiting eigenvalue distributions of the
sum to be given in a rather general and simple form. From the point of view
of the random matrix theory the problem that we
are going to consider is a version of the problem of the
deformation (see e.g. \cite{Br:81} for this term) of a given random matrix
(that can be a non-random matrix in particular) by another random matrix in
the case when ''randomness'' of the latter includes as an independent part
the random choice of the basis in which this matrix is diagonal. We will
discuss this topic in more details in Section 2.

In this paper we present a simple method of deriving functional equations
for the limiting eigenvalue distribution in a rather general situation. The
method is based on certain differential identities for expectations of
smooth matrix functions with respect to the normalized Haar measure of $U(n)$
( or $O(n)$ ) and on elementary matrix identities, the resolvent identity
first of all. The basic idea is the same as in \cite{Ma-Pa:67,Pa:72}: to
study not the moments of the counting measure, as it was proposed in the
pioneering paper by Wigner \cite{Wi:58}, but rather its Stieltjes (called
also the Cauchy or the Borel) transform, playing the role of appropriate
generating (or characteristic) function of the moments (the measure).
However, the technical implementation of the idea in this paper is different
and simpler then in \cite{Ma-Pa:67,Pa:72} (see Remark \ref{rem:1}
after Theorem \ref{th:main}).

The paper is organized as follows. In Section 2 we present our main results
(Theorem \ref{th:main})
and give their discussion. In Section 3 we prove Theorems \ref{th:3.1}
and \ref{th:3.2}
giving the solution of the problem under the conditions of the
uniform in $n$ boundedness of the forth moments of the normalized counting
measure of the terms. These conditions are more restrictive than those for
our principle result, given in Theorem \ref{th:main}.
Their advantage is that they
allow us to use the main ingredients of our approach in more transparent and
free of technicalities form. In Section 4 we prove Theorem \ref{th:main},
whose main
condition is the uniform boundedness of the first absolute moment of the
normalized counting measure of one of the two terms of the sum. In Section 5
we study certain properties of solutions of the functional equation and of
the limiting counting measure. In Section 6 we discuss topics related to our
main result and our technique.

\setcounter{equation}{0}
\setcounter{theorem}{0}
\setcounter{proposition}{0}
\setcounter{lemma}{0}
\setcounter{remark}{0}

\section{Model and Main Result.}

We consider the ensemble of $n$-dimensional Hermitian (or real symmetric)
random matrices $H_{n}$ of the form
\begin{equation}
H_{n}=H_{1,n}+H_{2,n},  \label{2.1}
\end{equation}
where
\[
H_{1,n}=V_{n}^{\ast }A_{n}V_{n},\ H_{2,n}=U_{n}^{\ast }B_{n}U_{n}.
\]
We assume that $A_{n}$ and $B_{n}$ are random Hermitian (or real symmetric)
matrices having arbitrary distributions, $V_{n}$ and $U_{n}$ are unitary (or
orthogonal) random matrices uniformly distributed over the unitary group $%
U(n)$ (or over the orthogonal group $O(n)$) with respect to the Haar
measure, and $A_{n}$, $B_{n}$, $%
V_{n}$ and $U_{n}$ are mutually independent. For the sake of definiteness we
will restrict ourself to the case of Hermitian matrices and the group $U(n)$
respectively. The results for symmetric matrices and for the group $O(n)$%
 have the same form, although their proof is more involved technically (see
Section 6).

We are interested in the asymptotic behavior as $n\rightarrow \infty $ of
the \textit{normalized eigenvalue counting measure} (NCM) $N_{n}$ of the
ensemble (\ref{2.1}), defined for any Borel set $\Delta \subset \mathbb{R}$%
\textbf{\ }by the formula
\begin{equation}
N_{n}(\lambda )=\frac{\#\{\lambda _{i}\in \Delta \}}{n},  \label{NCM}
\end{equation}
where $\lambda _{i},i=1,...,n$ are the eigenvalues of $H_{n}$.

The problem was studied recently \cite{Vo-Dy-Ni:92,Sp:93,Vo:98} in the context of
free (non-commutative) probability. In particular, it follows from results
of \cite{Sp:93} that if the matrices $A_{n}$ and $B_{n}$ are non-random,
their norms are uniformly bounded in $n$, i.e. their NCM $N_{1,n}$ and $%
N_{2,n}$ have uniformly in $n$ compact supports and if these measures have
weak limits as $n\rightarrow \infty $
\begin{equation}
N_{1,n}\rightarrow N_{1},\ N_{2,n}\rightarrow N_{2},  \label{N12}
\end{equation}
then the NCM (\ref{NCM}) of random matrix (\ref{2.1}) converges weakly with
probability 1 to a non-random measure $N$. Besides, if
\begin{equation}
f(z)=\int_{-\infty }^{\infty }\frac{N(\mathrm{d}\lambda )}{\lambda -z},
\quad \mathrm{Im}z > 0,
\label{st}
\end{equation}
is the \textit{Stieltjes transform} of this limiting measure and
\begin{equation}
f_{r}(z)=\int_{-\infty }^{\infty }\frac{N_{r}(\mathrm{d}\lambda )}{\lambda -z%
},\quad r=1,2,  \label{f12}
\end{equation}
are the Stieltjes transforms of $N_{r},$ \ $r=1,2$ of (\ref{N12}), then
according to \cite{Ne-Sp:95} $f(z)$ satisfies the functional equation
\begin{equation}
f(z)=f_{1}(z+R_{2}(f(z))),  \label{fl}
\end{equation}
where $R_{2}(f)$ is defined by the relation
\begin{equation}
z=-\frac{1}{f_2(z)}-R_{2}(f_2(z)))  \label{Rt}
\end{equation}
and is known as $R$-transform of the measure $N_{2}$ of (\ref{N12}) (see
Remark \ref{rem:3} after Theorem \ref{th:main}
and \cite{Vo-Dy-Ni:92,Vo:97} for the definition
and properties of this transform taking into account that our definition (%
\ref{Rt}) differs from that of \cite{Vo-Dy-Ni:92} by the sign). The proof of
this result in \cite{Sp:93,Ne-Sp:95} was based on the asymptotic analysis
of the expectations $m_{k}^{(n)}$ of moments of measure (\ref{NCM}). Since,
according to the spectral theorem and the definition (\ref{NCM}),
\begin{equation}
m_{k}^{(n)}=\mathbf{E}\{M_{k}^{(n)}\},\ M_{k}^{(n)}=n^{-1}\mathrm{Tr}%
H_{n}^{k},  \label{mkn}
\end{equation}
one can study the averaged moments $m_{k}^{(n)}$ by computing asymptotically
the expectations of the divided by $n$ traces of the $k$-th powers of (\ref
{2.1}), i.e. of corresponding multiple sums. This direct method dates back
to the classic paper by Wigner \cite{Wi:58} and requires a considerable
amount of combinatorial analysis, existence of all moments measures $%
N_{1,2}^{(n)}$ and their rather regular behavior as $n\rightarrow \infty $
to obtain the convergence of expectations (\ref{mkn}) for all integer $k$
and to guarantee that limiting moments determine uniquely corresponding
measure. By using this method it was proved in \cite{Sp:93,Ne-Sp:95} that
the expectation of $N_{n}$ converges to the limit, determined by (\ref{fl})
- (\ref{Rt}) and in \cite{Sp:93} that the variance $\mathrm{Var}%
\{M_{k}^{(n)}\}=$ $\mathbf{E}\{(M_{k}^{(n)})^{2}\}-\mathbf{E}%
^{2}\{M_{k}^{(n)}\}$ admits the bound
\begin{equation}
\mathrm{Var}\{M_{k}^{(n)}\}\leq \frac{C_{k}}{n^{2}},  \label{vmkn}
\end{equation}
where $C_{k}$ is independent of $n$. This bound yields evidently the
convergence of all moments with probability 1, thereby the weak convergence
with probability 1 of random measures (\ref{NCM}) to the non-random limit,
determined by (\ref{fl}) - (\ref{Rt}). The convergence with probability 1
here and below is understood as that in the natural probability space
\begin{equation}
\Omega =\prod_{n}\Omega _{n},  \label{Om}
\end{equation}
where $\Omega _{n}$ is the probability space of matrices (\ref{2.1}) that is
the product of respective spaces of $A_{n}$ and $B_{n}$ and two copies of
the group $U(n)$ for $U_{n}$ and $V_{n}$.

In this paper we obtain the analogous result under weaker assumptions and by
using a method, that does not involve combinatorics. This is because we work
with the Stieltjes transforms of measures (\ref{NCM}) and (\ref{N12}) and
derive directly the functional equations for their limits and the bound
analogous to (\ref{vmkn}) for the rate of their convergence (rather well
known in the random matrix theory, see e.g. \cite{Pa-Sh:97,KKP:96}) by
using certain simple identities for expectations of matrix functions with
respect to the Haar measure (Proposition \ref{pr:3.2}
below) and elementary facts on
resolvents of Hermitian matrices.

The Stieltjes transform was first used in studies of the
eigenvalue distribution of random matrices in paper \cite{Ma-Pa:67}
and proved to be an efficient tool in the field (see e.g.
\cite{Gi:75,Gi:90,KKP:96,KKPS:92,Kh:96,Pa-Kh:93,Pa:72,Pa:96,Pa:99,Silv:95,Si-Ch:95}).
We list the properties of the
Stieltjes transform that we will need below (see e.g.\cite{Ak-Gl:63}).

\begin{proposition}
\label{pr:2.1} Let $m$ be a
non-negative and normalized to unity measure and
\begin{equation}
s(z)=\int \frac{m(d\lambda )}{\lambda -z},\quad \mathrm{Im\,~}z \neq 0  \label{St}
\end{equation}
 be the Stieltjes transform of $m$ (here and below integrals
without limits denote the integrals over the whole axis). Then:

\begin{enumerate}
\item[(i)]  $s(z)$ is analytic in $\mathbb{C}\setminus \mathbb{R}$
and
\begin{equation}
|s(z)|\leq |\mathrm{Im\,~}z|^{-1};  \label{bound}
\end{equation}
\item[(ii)]
\begin{equation}
\mathrm{Im\,~}s(z)\mathrm{Im\,~}z>0,\ \mathrm{Im\,~}z\neq 0;  \label{Nev}
\end{equation}
\item[(iii)]
\begin{equation}
\lim_{y\rightarrow \infty }y|s(iy)|=1;  \label{as}
\end{equation}
\item[(iv)]   for any continuous function $\varphi $ with
a compact support we have the inversion (Frobenius-Perron) formula
\begin{equation}
\int \phi (\lambda )N(\mathrm{d}\lambda )=\lim_{\varepsilon \rightarrow 0}%
\frac{1}{\pi }\int \phi (\lambda )\mathrm{Im\,~}s(\lambda +i\varepsilon );
\label{FP}
\end{equation}

\item[(v)]   conversely, any function verifying (\ref{bound}) - (%
\ref{as}) is the Stieltjes transform of a non-negative and normalized to
unity measure and this one-to-one correspondence between measures and their
Stieltjes transforms is continuous if one will use the topology of weak
convergence for measures and the topology of convergence on compact sets of
$\mathbb{C}\setminus \mathbb{R}$\textit{\ for their Stieltjes transforms.}
\end{enumerate}
\end{proposition}

We formulate now our main result. Since eigenvalues of a Hermitian matrix
are unitary invariant we can replace matrices (\ref{2.1}) by
\begin{equation}
H_{n}=A_{n}+U_{n}^{\ast }B_{n}U_{n},  \label{ABU}
\end{equation}
where $A_{n}$, $B_{n}$ and $U_{n}$ are the same as in (\ref{2.1}). However,
it is useful to keep in mind that the problem is symmetric in $A_{n}$ and $%
B_{n}$. We prove

\begin{theorem}
  \label{th:main}
Let $H_{n}$ be the random $n\times
n $ matrix of the form (\ref{2.1}). Assume that the normalized eigenvalue
counting measures $N_{r,n},r=1,2$ of matrices $A_{n}$ and $B_{n}$ converge
weakly in probability as $n\rightarrow \infty $ to the non-random
nonnegative and normalized to 1 measures $N_{r},r=1,2$ respectively and that
\begin{equation}
\sup_{n}\int |\lambda |\mathbf{E}N_{r,n}^{\ast }(\mathrm{d}\lambda )\leq
m_{1}<\infty ,  \label{2.2}
\end{equation}
where $N_{r,n}^{\ast }$ is one of the measures $N_{1,n}$ or $N_{2,n}$. Then
the normalized eigenvalue counting measure $N_{n}$of $H_{n}$ converges in
probability to a non-random nonnegative and normalized to 1 measure $N$
whose Stieltjes transform (\ref{st}) is a unique solution of the system
\[
f(z)=f_{1}\left( z-{\frac{\Delta _{2}(z)}{f(z)}}\right)
\]
\begin{equation}
f(z)=f_{2}\left( z-{\frac{\Delta _{1}(z)}{f(z)}}\right)  \label{2.4}
\end{equation}
\[
f(z)={\frac{1-\Delta _{1}(z)-\Delta _{2}(z)}{-z}}
\]
in the class of functions $f(z)$ satisfying (\ref{bound}) - (\ref{as}) and
functions $\Delta _{r}(z),r=1,2$ analytic for $\mathrm{Im\,~}z\neq 0$ and
satisfying conditions
\begin{equation}
\Delta _{1,2}(z)\rightarrow 0\ \mathrm{as}\ \mathrm{Im\,~}z\rightarrow
\infty ,  \label{2.6}
\end{equation}
where $f_{r}(z),r=1,2$ are Stieltjes transforms (\ref{f12}) of the measures $%
N_{r},r=1,2$ and $\mathbf{E}\{\cdot \}$ denotes the expectation with respect
to the probability measure, generated by $A_{n}$, $B_{n}$, $U_{n}$ and $%
V_{n}$
\end{theorem}

The theorem will be proved in Section 4. Here we make several remarks
related to the theorem (see also Section 5).

\begin{remark}
\label{rem:1}{\rm The historically first example of a random matrix
ensemble representable in the form (\ref{ABU}) was proposed in \cite
{Ma-Pa:67} and has the form
\begin{equation}
H_{m,n}=H_{0,n}+\sum\limits_{i=1}^{m}\tau _{i}P_{q_{i}},  \label{MP}
\end{equation}
where $H_{0,n}$ is a non-random $n\times n$ Hermitian matrix such that its
normalized eigenvalue counting measure converges weakly to a limiting
nonnegative and normalized to 1 measure $N_{0}$, $\tau _{i}$, $i=1,..m$ are
i.i.d. random variables and $P_{q_{i}}$ are orthogonal projections on unit
vectors $q_{i}$, $i=1,..m$ uniformly and independently of one another and of
$\{\tau _{i}\}_{i=1}^{m}$ distributed over the unit sphere in $\mathbb{C}%
^{n} $ {\footnote{%
In fact, in \cite{Ma-Pa:67} a more general class of independent random
vectors was considered, but we restrict ourself here to the unit vectors, in
order to have an example of an ensemble of form (\ref{2.1}).}}. It is clear
that the matrix
\begin{equation}
\sum\limits_{i=1}^{m}\tau _{i}P_{q_{i}}  \label{MPO}
\end{equation}
can be written in the form $U_{n}^{\ast }B_{n}U_{n}$ of the second term of (%
\ref{2.1}) or (\ref{ABU}). According to \cite{Ma-Pa:67} the NCM of random
matrix (\ref{MPO}) converges in probability as $n\rightarrow \infty $, $%
m\rightarrow \infty $, $m/n\rightarrow c\geq 0$ to a non-random nonnegative
and normalized to 1 measure whose Stieltjes transform $f_{MP}(z)$ satisfies
the equation
\begin{equation}
f_{MP}(z)=-\left( z+c\int \frac{\tau \sigma (\mathrm{d}\tau )}{1+\tau
f_{MP}(z)}\right) ^{-1},  \label{fMPO}
\end{equation}
where $\sigma $ is the probability law of $\tau _{i}$ in (\ref{MP}). Assume
that $\sigma $ has the finite first moment
\begin{equation}
\int |\tau |\sigma (\mathrm{d}\tau )<\infty  \label{tau}
\end{equation}
Then taking (\ref{MPO}) as the second term of (\ref{2.1}) we get, in view of
inequality
\[
\mathbf{E}\left\{ \int |\lambda |N_{2,n}(\mathrm{d}\lambda )\right\} \leq
n^{-1}\sum\limits_{i=1}^{m}\mathbf{E}\{|\tau _{i}|\}=\frac{m}{n}\mathbf{E}%
\{|\tau |\}<\infty ,
\]
the condition (\ref{2.2}) of Theorem \ref{th:main}.
Applying then Theorem \ref{th:main} in which
$f_{2}(z)$ is given by (\ref{fMPO}), we obtain from the two last equations
of the system (\ref{2.4}) that
\[
\frac{\Delta _{1}(z)}{f(z)}=c\int \frac{\tau \sigma (\mathrm{d}\tau )}{%
1+\tau f_{MP}(z)}.
\]
This and the first equation of (\ref{2.4}) yield the functional equation for
the Stieltjes transform of the limiting eigenvalue distribution of ensemble (%
\ref{MP})
\begin{equation}
f(z)=f_{0}\left( z-c\int \frac{\tau \sigma (\mathrm{d}\tau )}{1+\tau f(z)}%
\right)  \label{fMP}
\end{equation}
where $f_{0}(z)$ is the Stieltjes transform of the limiting NCM $N_{0}$ of
the non-random matrix $H_{0,n}$. This equation was obtained in \cite
{Ma-Pa:67} by another method, whose main ingredient was careful analysis of
changes of the resolvent of matrices (\ref{MP}) induced by addition of the $%
(m+1)$-th term, i.e. by a rank-one perturbation. This allowed the authors to
prove that the sequence $g_{i,n}(z)=n^{-1}\mathrm{Tr}(H_{i,n}-z)^{-1},\
i=1,...,m$ converges in probability to the non-random limit $f(z,t),\,z\in
\mathbb{C}\backslash \mathbf{R}, \, t\in \lbrack 0,1]$, as $n\rightarrow \infty
,m\rightarrow \infty ,\,i\rightarrow \infty ,\,m/n\rightarrow
c,\,i/m\rightarrow t$, and that the limiting function $f(z,t)$ satisfies the
quasilinear PDE
\begin{equation}
\frac{\partial f}{\partial t}+c\frac{\tau (t)}{1+\tau (t)f}\frac{\partial f}{%
\partial z},\;f(z,0)=f_{0}(z),  \label{Burg}
\end{equation}
where $\tau (t)$ is the inverse of the probability distribution $\sigma
(\tau )=\mathbf{P}\{\tau _{i}\leq \tau \}$. It can be shown that the
solution of (\ref{Burg}) at $t=1$ coincides with (\ref{MP}) \cite{Ma-Pa:67}.
Equation (\ref{Burg}) with $\tau (t)\equiv \mathrm{const}$ is a particular
case of the so-called complex Burgers equation appeared in the free
probability \cite{Vo-Dy-Ni:92}, where the random matrices (\ref{MP}) provide
an analytic model for the stationary processes with free increments, like in
the conventional probability the heat equation and sums of i.i.d. random
variables comprise an important ingredient of the theory of random
processes with independent increments.}
\end{remark}

\begin{remark}
\label{rem:2}{\rm Consider the ensemble known as the deformed Gaussian
ensemble \cite{Pa:72}:
\begin{equation}
H_{n}=H_{0,n}+M_{n},  \label{dGUE}
\end{equation}
where $H_{0,n}$ is a non-random matrix such that its normalized eigenvalue
counting measure converges weakly to the limit $N_{0}$ and $%
M_{n}=\{M_{jk}\}_{j,k=1}^{n}$ is a random Hermitian matrix whose matrix
elements $M_{jk}$ are complex Gaussian random variables satisfying
conditions:
\begin{equation}
\overline{M_{jk}}=M_{kj},\ \mathbf{E}\{M_{jk}\}=0,\ \mathbf{E}%
\{M_{j_{1}k_{1}}\overline{M_{j_{2}k_{2}}}\}=\frac{2w^{2}}{n}\delta
_{j_{1}j_{2}}\delta _{k_{1}k_{2}}.  \label{GUEm}
\end{equation}
In other words, the ensemble is defined by the distribution
\begin{equation}
\mathbf{P}(\mathrm{d}M)=Z_{n}^{-1}exp\left\{ -\frac{n}{4w^{2}}\mathrm{Tr}%
M^{2}\right\} \mathrm{d}M,\   \label{GUE}
\end{equation}
\[
\mathrm{d}M=\prod\limits_{j=1}^{n}\mathrm{d}M_{jj}\prod\limits_{1\leq
j<k\leq n}\mathrm{d}\mathrm{Re\,}M_{jk}\mathrm{dIm\,}M_{jk},
\]
where $Z_{n}$ is the normalization constant. The distribution defines the
Gaussian Unitary Ensemble (GUE) \cite{Me:91}. This is why ensemble (\ref
{dGUE}) is called the deformed GUE \cite{Br:81}. It is known \cite{Me:91}
that $M_{n}$ can be written in the form
\begin{equation}
M_{n}=U_{n}^{\ast }\Lambda _{n}U_{n},  \label{MGUE}
\end{equation}
where $U_{n}$ are unitary matrices whose probability law is the Haar measure
on $\mathit{U}(n)$ and $\Lambda _{n}$ is independent of $U_{n}$ diagonal
random matrix whose normalized eigenvalue counting measure converges with
probability 1 to the semicircle law. The Stieltjes transform $f_{sc}(z)$ of
the latter satisfies the simple functional equation \cite{Pa:72}
\begin{equation}
f_{sc}(z)=-(z+2w^{2}f_{sc}(z)),  \label{sc}
\end{equation}
whose solution yields the semicircle law by Wigner
\begin{equation}
N_{sc}(d\lambda )=(4\pi w^{2})^{-1}\sqrt{8w^{2}-\lambda ^{2}}\chi _{\lbrack
-2\sqrt{2}w,2\sqrt{2}w]}(\lambda )d\lambda ,  \label{Wi}
\end{equation}
where $\chi _{\lbrack a,b]}(\lambda )$ is the indicator of the interval $%
[a,b]\subset \mathbb{R}$. It is easy to see that
\[
\mathbf{E}\{n^{-1}\mathrm{Tr}M_{n}^{2}\}=2w^{2}<\infty .
\]
Denoting $N_{sc,n}$ the NCM of the random matrices defined by (\ref{GUE}) we
can rewrite this inequality in the form
\begin{equation}
\int_{-\infty }^{\infty }\lambda ^{2}\mathbf{E}\{N_{sc,n}(d\lambda
)\}<\infty .  \label{2.281}
\end{equation}
Thus, if we use (\ref{MGUE}) as the second term in (\ref{ABU}), it will
satisfy condition (\ref{2.1}). Taking $\ f_{sc}(z)$ as $f_{2}(z)$ in (\ref
{2.4}) we find from the two last equations of the system that $\Delta
_{2}(z)/f(z)=-2w^{2}f(z)$ and then the first equation of (\ref{2.4}) takes
the form
\begin{equation}
f(z)=f_{0}(z+2w^{2}f(z)),  \label{dsc}
\end{equation}
where $f_{0}(z)$ is the Stieltjes transform of the limiting counting measure
of matrices $H_{0,n}$. This functional equation determining the limiting
eigenvalue distribution of the deformed GUE was found by another method in
\cite{Pa:72} (see also \cite{KKPS:92}) for random matrices (\ref{dGUE}) in
which $M_{n}$ has independent (modulo the Hermitian symmetry conditions)
entries, for (\ref{GUE}) in particular.
}\end{remark}

\begin{remark}
\label{rem:3}
{\rm Consider now a probability measure $m(\mathrm{d}\lambda )$ and assume
that its second moment $m_{2}$ is finite. In this case we can write the
Stieltjes transform $s(z)$ of $m$ in the form
\[
s(z)=-(z+\Sigma (z))^{-1},
\]
where $\Sigma (z)$ is the Stieltjes transform of a non-negative measure
whose total mass is $m_{2}$ (to prove this fact one can use, for example,
the general integral representation \cite{Ak-Gl:63} for functions satisfying
(\ref{Nev}) ). Since $s^{\prime }(z)=z^{-2}(1+o(1))$, $z\rightarrow \infty ,$
then, according to the local inversion theorem, there exists a unique
functional inverse $z(s)$ of $s(z)$ defined and analytic in a neighborhood
of zero and assuming its values in a neighborhood of infinity. Denote
\begin{equation}
\Sigma (z(s))=R_{m}(s)  \label{sigR}
\end{equation}
and following Voiculescu \cite{Vo-Dy-Ni:92} call $R_{m}(s)$ the $R$%
-transform of the probability measure $m$. By using the $R$-transforms $%
R_{1,2}$ of measures $N_{1,2}$ we can rewrite the first two equations of
system (\ref{2.4}) in the form
\begin{equation}
\frac{\Delta _{1,2}}{f(z)}=\frac{1}{f(z)}%
+z+R_{2,1}(f(z))=-R(f(z))+R_{2,1}(f(z)),  \label{R12}
\end{equation}
where $R$ denotes the $R$-transform of the limiting normalized counting measure $%
N$ of the ensemble (\ref{2.1}) (the measure whose Stieltjes transform is $%
f$). These relations and  the third
equation of system (\ref{2.4}) lead to the remarkably simple expression of $R$ via $%
R_{1}$ and $R_{2}$
\begin{equation}
R(f)=R_{1}(f)+R_{2}(f),  \label{R}
\end{equation}
that ''linearizes'' the rather complex system (\ref{2.4}). The relation was
obtained by Voiculescu in the context of $C^{\ast }$-algebra studies (see
\cite{Vo-Dy-Ni:92,Vo:97} for results and references). Thus, one can regard
the system (\ref{2.4}) as a version of the binary operation on measures
defined by (\ref{R}) and known as the non-commutative convolution. A simple
precursor of relation (\ref{R}) containing the functional inverses of $f$
and $f_{1,2}$ for real $z$ lying outside of the support of $N_{0}$ in (\ref
{fMP}) was used in \cite{Ma-Pa:67} (see also \cite{Si-Ch:95}) to locate the
support of $N$ in terms of the support of $N_{0}$ in the case of ensemble (%
\ref{MP}). The simplest form of the relation (\ref{R}) for the case when
both measures are semicircle measures (\ref{Wi}), i.e. when $%
R_{1,2}=2w_{1,2}^{2}f$, was indicated in \cite{Pa:72}. Formal derivation of
relation (\ref{R}) for the case then matrices $H_{1}$ and $H_{2}$
distributed both according to the laws
\begin{equation}
P_{1,2}^{(n)}(\mathrm{d}H)=Z_{1,2}^{(n)}exp\{-nV_{1,2}(H)\}\mathrm{d}H.
\label{MM}
\end{equation}
where $V_{1,2}:\mathbb{R}\rightarrow \mathbb{R}_{+}$ are polynomials of an
even degree was given in \cite{Zee:96}. The derivation is based on the
perturbation theory with respect to the non-quadratic part of $V_{1,2}$ and
the $R$-transform is related to the sum of irreducible diagrams of the
formal perturbation series. Existence of the limiting eigenvalue counting
measure for the random matrix ensemble (\ref{MM}) was rigorously proved in
\cite{ABM-Pa-Sh:95} for a rather broad class of functions $V$ (not necessary
polynomials). It was also proved that the normalized counting measure (\ref
{NCM}) converges in probability to the limiting measure. The form (\ref{MGUE}%
) of matrices of ensemble (\ref{MM}) can be deduced from known results on
the ensemble (\ref{MM}) (see e.g.\cite{Be-It-Zu:80}) in the same way as for
the GUE (\ref{GUE}), where $V(\lambda )=\lambda ^{2}/4w^{2}$ (see \cite
{Me:91}). Condition (\ref{2.2}) follows from results of
\cite{ABM-Pa-Sh:95,Pa:99}.
Thus we can apply Theorem \ref{th:main} to obtain rigorously relation
(\ref{R})
in the case when matrices $H_{r},r=1,2$ in (\ref{2.1}) are distributed
according to (\ref{MM}).
}\end{remark}

\begin{remark}
{\rm The problem of addition of random Hermitian (real symmetric matrices)
has a natural multiplicative analogues in the case of positive
defined Hermitian (real symmetric) or unitary (orthogonal) matrices.
Namely, assuming that $A_n$ and $B_n$ are positive defined
matrices and $U_n$ is the unitary (orthogonal) Haar distributed
random matrix we can consider the positive defined random matrix
\begin{equation}
H_{n}=A_{n}^{1/2}U_{n}^{\ast }B_{n}U_{n}A_{n}^{1/2}.  \label{2.2.1}
\end{equation}
Likewise, if $S_n$ and $T_n$ are unitary (orthogonal) matrices
and and $U_n$ is as above we can consider the random matrices
\begin{equation}
V_n=S_{n}U_{n}^{\ast }T_{n}U_{n}.
\label{2.1.1}
\end{equation}
In this case the normalized eigenvalue counting measure is defined
as $n^{-1}$ times the number of eigenvalues belonging to a Borel
set of the unit circle.

In both cases (\ref{2.2.1}) and (\ref{2.1.1}) one can study the
limiting properties of the NCM's of respective random matrices
provided that the "input" matrices $A_n,B_n,S_n$
and $T_n$ have limiting eigenvalue distributions. The first
examples of ensembles of the above forms as multiplicative analogues
of the ensemble (\ref{MP}) were proposed in \cite{Ma-Pa:67},
where the respective functional equations analogous to (\ref{fMP}) were
derived. A general class of the random matrix ensembles of these
forms were studied in free probability
\cite{Vo:91,Vo-Dy-Ni:92,Be-Vo:93}, where the notions
of the $S$ - transform and the free multplicative convolution of
measures were
proposed and used to give a general form of the limiting
eigenvalue distributions of products (\ref{2.2.1}) and (\ref{2.1.1}).
It will be shown in the subsequent paper \cite{Va:00} that a
version of the method of this paper leads to results, analogous
to those given in Theorem 2.1 above.
}\end{remark}

\setcounter{equation}{0}
\setcounter{theorem}{0}
\setcounter{proposition}{0}
\setcounter{lemma}{0}
\setcounter{remark}{0}
\section{Convergence with Probability 1 for non-Random $A_n$ and $B_n$.}

As the first step of the proof of Theorem \ref{th:main}
we prove the following
\begin{theorem}
\label{th:3.1} Let $H_{n}$ be the random $n\times
n $ matrix of the form (\ref{2.1}) in which $A_{n}$ and $B_{n}$ are
non-random Hermitian matrices, $U_{n}$ and $V_{n}$ are random independent
unitary matrices distributed each according to the normalized to unity Haar
measure on $\mathit{U}(n)$. Assume that the normalized counting measures $%
N_{r,n},r=1,2$ of matrices $A_{n}$ and $B_{n}$ converge weakly as $%
n\rightarrow \infty $ to nonnegative and normalized to 1 measures $%
N_{r},r=1,2$ respectively and that
\begin{equation}
\sup_{n}\int \lambda ^{4}N_{r,n}(\mathrm{d}\lambda )=m_{4}<\infty ,r=1,2.
\label{3.1}
\end{equation}
Then the normalized eigenvalue counting measure (\ref{NCM}) of $H_{n}$
converges with probability 1 to a non-random and normalized to 1 measure
whose Stieltjes transform (\ref{st}) is a unique solution of the system
(\ref{2.4})
in the class of functions $f(z)$, $\Delta _{r}(z),r=1,2$ analytic for
$\mathrm{Im\,~}z\neq 0$ and satisfying conditions (\ref{bound})-(\ref{as})
and (\ref{2.6}) respectively.
\end{theorem}

\begin{remark} {\rm The theorem generalizes the results of \cite{Sp:93}
proved under the condition that supports of the NCM $N_{r,n},r=1,2$ of $%
A_{n} $ and $B_{n}$ are uniformly bounded in $n.$
}\end{remark}

\begin{remark}
{\rm By mimicking the proof of the Glivenko - Cantelli theorem (see e.g.
\cite{Lo:77}), one can prove that the random distribution functions $%
N_{n}(\lambda )=N_{n}(]-\infty ,\lambda [)$ corresponding to measures (\ref
{NCM}) converge uniformly with probability 1 to the distribution function $%
N(\lambda )=N(]-\infty ,\lambda [)$ corresponding to measure $N$:
\[
\mathbf{P}\{\lim_{n\rightarrow \infty }\sup_{\lambda \in \mathbb{R}%
}|N_{n}(\lambda )-N(\lambda )|=0\}=1.
\]
}\end{remark}

We present now our technical means. First is a collection of elementary
facts of linear algebra.

\begin{proposition}
\label{pr:3.1} Let $\mathbf{M}_{n}$ be the
algebra of linear transformations of $\mathbb{C}^{n}$ in itself ($n\times n$
complex matrices) equipped with the norm, induced by the Euclidean norm of $%
\mathbb{C}^{n}$.

We have :
\begin{enumerate}
\item[(i)]  if $M\in \mathbf{M}_{n}$ and $\{M_{jk}\}_{j,k=1}^{n}$ is
the matrix of $M$ in any orthonormalized basis of $\mathbb{C}^{n}$, then
\begin{equation}
|M_{jk}|\leq ||M||;  \label{3.3}
\end{equation}

\item[(ii)]  if $\mathrm{Tr}M=\sum\limits_{j=1}^{n}M_{jj}$, then
\begin{equation}
|\mathrm{Tr}M_{1}M_{2}|\leq (\mathrm{Tr}M_{1}M_{1}^{\ast })^{1/2}(\mathrm{Tr}%
M_{2}M_{2}^{\ast })^{1/2},  \label{3.4}
\end{equation}
where $M^{\ast }$ is the Hermitian conjugate of $M$, and if $P$ is a
positive defined transformation, then
\begin{equation}
|\mathrm{Tr}MP|\leq ||M||\mathrm{Tr}P;  \label{3.5}
\end{equation}

\item[(iii)]  for any Hermitian transformation $M$ its {resolvent}
\begin{equation}
G(z)=(M-z)^{-1}  \label{3.6}
\end{equation}
is defined for all non-real $z$, $\mathrm{Im\,~}z\neq 0$,
\begin{equation}
||G(z)||\leq |\mathrm{Im\,~}z|^{-1}  \label{3.7}
\end{equation}
and if $\{G_{jk}(z)\}_{j,k=1}^{n}$ is the matrix of $G(z)$ in any
orthonormalized basis of $\mathbb{C}^{n}$ then
\begin{equation}
|G_{jk}(z)|\leq |\mathrm{Im\,~}z|^{-1};  \label{3.8}
\end{equation}

\item[(iv)]  if $M_{1}$ and $M_{2}$ are two Hermitian
transformations and $G_{r}(z),r=1,2$ are their resolvents, then
\begin{equation}
G_{2}(z)=G_{1}(z)-G_{1}(z)(M_{2}-M_{1})G_{2}(z)  \label{3.9}
\end{equation}
(the resolvent identity);

\item[(v)]  if $G(z)=(M-z)^{-1}$ is regarded as a function of $M,$
then the derivative $G^{\prime }(z)$ of $G(z)$ with respect to $M$
verifies the relation
\begin{equation}
G^{\prime }(z)\cdot X=-G(z)XG(z)  \label{3.10}
\end{equation}
for any Hermitian $X\in \mathbf{M}_{n}$, and, in particular,
\begin{equation}
||G^{\prime }(z)||\leq ||G(z)||^{2}\leq |\mathrm{Im\,~}z|^{-2}  \label{3.11}
\end{equation}
\end{enumerate}
\end{proposition}

Now is our main technical tool.

\begin{proposition}
\label{pr:3.2} Let $\Phi :\mathbf{M}_{n}\rightarrow \mathbb{C}$
be a continuously differentiable function. Then
the following relation holds for any $M\in \mathbf{M}_{n}$ and any Hermitian
element $X\in \mathbf{M}_{n}$:
\begin{equation}
\int\limits_{U(n)}\Phi ^{\prime }(U^{\ast }MU)\cdot \lbrack X,U^{\ast }MU]%
\mathrm{d}U=0,  \label{3.12}
\end{equation}
where
\begin{equation}
\mathit{\ [M_{1},M_{2}]=M_{1}M_{2}-M_{1}M_{2}}  \label{com}
\end{equation}
is the commutator of $\mathit{M_{1}}$ and \, $\mathit{M_{2}}$%
and the symbol
\begin{equation}
\int\limits_{U(n)}\mathit{...dU}  \label{int}
\end{equation}
denotes the integration over $\mathit{U}(n)$ with
respect to the normalized Haar measure $\mathit{dU}$.
\end{proposition}
\medskip

\textit{Proof.} To prove (\ref{3.12}) we use the right shift invariance of
the Haar measure: $\mathrm{d}U=\mathrm{d}(UU_{0})$, $\forall U_{0}\in U(n)$
according to which the integral
\[
\int\limits_{U(n)}\Phi \left( e^{-i\varepsilon X}U^{\ast }MUe^{i\varepsilon
X}\right) \mathrm{d}U
\]
is independent of $\varepsilon $ for any Hermitian $X\in \mathbf{M}_{n}$.
Thus its derivative with respect to $\varepsilon $ at $\varepsilon =0$ is
zero. This derivative is the l.h.s. of (\ref{3.12}).$\blacksquare $

\begin{proposition}

\label{pr:3.3} System (\ref{2.4}) has a unique
solution in the class of functions $f(z)$, $\Delta _{1,2}(z)$ analytic for $%
\mathrm{Im\,~}z\neq 0$ and satisfying conditions (\ref{bound})--(\ref{as})
and (\ref{2.6}).
\end{proposition}
 \medskip

\textit{Proof.} Assume that there exist two solutions $(f^{^{\prime
}},\Delta _{1,2}^{^{\prime }})$ and $(f^{^{\prime \prime }},\Delta
_{1,2}^{^{\prime \prime }})$ of the system. Denote $\delta f=f^{^{\prime
}}-f^{^{\prime \prime }}$, $\delta \Delta _{1,2}=\Delta _{1,2}^{^{\prime
}}-\Delta _{1,2}^{^{\prime \prime }}$. Then, by using (\ref{2.4}) and the
integral representation (\ref{f12}) for $f_{1,2},$ we obtain the linear
system for $\delta \phi =z\delta f$, and for $\delta \Delta _{1,2}$
\begin{equation}
\begin{array}{ll}
\delta \phi (1-a_{1}(z))+b_{1}(z)\delta \Delta _{1} & =0, \\
\delta \phi (1-a_{2}(z))+b_{2}(z)\delta \Delta _{2} & =0, \\
\delta \phi -\delta \Delta _{1}-\delta \Delta _{2} & =0,
\end{array}
\label{ls}
\end{equation}
where
\begin{equation}
a_{1}=\frac{\Delta _{1}^{^{\prime \prime }}}{f^{^{\prime }}f^{^{\prime
\prime }}}I_{2},\ b_{1}=\frac{z}{f^{^{\prime }}}I_{2},\ I_{2}=I_{2}(z-\Delta
_{1}^{^{\prime }}/f^{^{\prime }},z-\Delta _{1}^{^{\prime \prime }}/f^{\prime
\prime }),  \label{3.12.1}
\end{equation}
\begin{equation}
I_{2}(z^{\prime },z^{\prime \prime })=\int \frac{N_{2}(\mathrm{d}\lambda )}{%
(\lambda -z^{\prime })(\lambda -z^{\prime \prime })},  \label{3.12.2}
\end{equation}
and $a_{2}$, $b_{2}$ can be obtained from $a_{1}$ and $b_{1}$ by replacing $%
N_{2}$ and $\Delta _{1}$ by $N_{1}$ and $\Delta _{2}$ in above formulas. For
any $y_{0}>0$ consider the domain
\begin{equation}
E(y_{0})=\{z\in \mathbb{C}:|\mathrm{Im\,~}z|\geq y_{0},\ |\mathrm{Re\,~}%
z|\leq |\mathrm{Im\,~}z|\}.  \label{E}
\end{equation}
If $s(z)$ is the Stieltjes transform (\ref{St}) of a probability measure $m,$
then we have for $z\in E(y_{0})$,
\[
\left| \int \frac{\lambda m(\mathrm{d}\lambda )}{\lambda -z}\right| =\left|
\int\limits_{|\lambda |\leq M}+\int\limits_{|\lambda |>M}\right| \leq \frac{M%
}{y_{0}}+2\int\limits_{|\lambda |>M}m(\mathrm{d}\lambda ),
\]
i.e.
\begin{equation}
zs(z)=-1+o(1),\ z\rightarrow \infty ,\ z\in E(y_{0}).  \label{3.3.1}
\end{equation}
Analogously, by using this asymptotic relation and condition (\ref{2.6}) we
obtain that for $z\rightarrow \infty $, $z\in E(y_{0})$
\[
z^{2}I_{1,2}(z)=1+o(1),\ a_{1,2}(z)=o(1),\ b_{1,2}(z)=-1+o(1).
\]
Thus the determinant $b_{1}b_{2}+b_{1}+b_{2}-(a_{2}b_{1}+a_{1}b_{2})$ of
system (\ref{ls}) is equal asymptotically to $-1$. We conclude that if $%
y_{0} $ in (\ref{E}) is big enough, then system (\ref{ls}) has only a
trivial solution, i.e. system (\ref{2.4}) is uniquely soluble. $\blacksquare
$\bigskip

In what follows we use the notation
\begin{equation}
\int\limits_{U(n)}...\mathrm{d}U=\langle ...\rangle  \label{3.13}
\end{equation}

\medskip
\textit{Proof of Theorem \ref{th:3.1}.}
Because of unitary invariance of eigenvalues
of a Hermitian matrices we can assume without loss of generality that the
unitary matrix $V$ in (\ref{2.1}) is set to unity, i.e. we can work with the
random matrix (\ref{ABU}). We will omit below the subindex $n$ in all cases
when it will not lead to confusion. Write the resolvent identity (\ref{3.9})
for the pair $(H_{1},H)$ of (\ref{2.1}):
\begin{equation}
G(z)=G_{1}(z)-G_{1}(z)H_{2}G(z),  \label{3.14}
\end{equation}
where
\[
G(z)=(H_{1}+H_{2}-z)^{-1},\ G_{1}(z)=(H_{1}-z)^{-1}.
\]
Consider the matrix $\langle g_{n}(z)G(z)\rangle ,$ where
\begin{equation}
g_{n}(z)={\frac{1}{n}}\mathrm{Tr}G(z)=\int \frac{N_{n}(\mathrm{d}\lambda )}{%
\lambda -z},\ \mathrm{Im\,~}z\neq 0  \label{3.15}
\end{equation}
is the Stieltjes transform of random measure (\ref{NCM}). The resolvent
identity (\ref{3.14}) leads to the relation
\begin{equation}
\langle g_{n}(z)G(z)\rangle =\langle g_{n}(z)\rangle
G_{1}(z)-G_{1}(z)\langle g_{n}(z)H_{2}G(z)\rangle .  \label{3.16}
\end{equation}
By using Proposition \ref{pr:3.2} with the matrix element $\left(
(H_{1}+M-z)^{-1}\right) _{ac}$ as $\Phi (M)$ we have in view of (\ref{3.10})
and (\ref{3.12}) - (\ref{com})
\[
\langle (G[X,H_{2}]G)_{ac}\rangle =0.
\]
Choosing the Hermitian matrix $X$ with only $(a,b)$-th and $(b,a)$ non-zero
entries, we obtain
\begin{equation}
\langle G_{aa}(H_{2}G)_{bc}\rangle =\langle (GH_{2})_{aa}G_{bc}\rangle .
\label{3.16.1}
\end{equation}
Applying to this relation the operation $n^{-1}\sum\limits_{a=1}^{n}$ and
taking into account the definition (\ref{3.15}) of $g_{n}(z)$ we rewrite the
last relation in the form
\[
\langle g_{n}(z)H_{2}G(z)\rangle =\langle \delta _{2,n}(z)G(z)\rangle ,
\]
where
\begin{equation}
\delta _{2,n}(z)={\frac{1}{n}}\mathrm{Tr}H_{2}G(z).  \label{3.17}
\end{equation}
Thus we can rewrite (\ref{3.16}) as
\begin{equation}
\langle g_{n}(z)G(z)\rangle =\langle g_{n}(z)\rangle
G_{1}(z)-G_{1}(z)\langle \delta _{2,n}(z)G(z)\rangle .  \label{3.18}
\end{equation}
Introduce now the centralized quantities
\begin{equation}
g_{n}^{\circ }(z)=g_{n}(z)-f_{n}(z),\ \delta _{2,n}^{\circ }(z)=\delta
_{2,n}(z)-\Delta _{2,n}(z),  \label{3.19}
\end{equation}
where
\begin{equation}
f_{n}(z)=\langle g_{n}(z)\rangle ,\ \Delta _{2,n}(z)=\langle \delta
_{2,n}(z)\rangle .  \label{3.20}
\end{equation}
With these notations (\ref{3.18}) becomes
\begin{equation}
f_{n}(z)\langle G(z)\rangle =f_{n}(z)G_{1}(z)-\Delta
_{2,n}(z)G_{1}(z)\langle G(z)\rangle +R_{1,n}(z),  \label{3.21}
\end{equation}
where
\begin{equation}
R_{1,n}(z)=-\langle g_{n}^{\circ }(z)G(z)\rangle -G_{1}(z)\langle \delta
_{2,n}^{\circ }(z)G(z)\rangle .  \label{R1n}
\end{equation}
Besides, since
\begin{equation}
\begin{array}{l}
{n^{-1}}\mathrm{Tr}H^{2}={n^{-1}}\mathrm{Tr}(H_{1}+H_{2})^{2}\leq 2{n^{-1}}%
\mathrm{Tr}H_{1}^{2}+2{n^{-1}}\mathrm{Tr}H_{2}^{2}= \\
=2\displaystyle\int \lambda ^{2}N_{1,n}(\mathrm{d}\lambda )+2\displaystyle%
\int \lambda ^{2}N_{2,n}(\mathrm{d}\lambda )\leq 4m_{2}\leq 4m_{4}^{1/2},
\end{array}
\label{TrH2}
\end{equation}
we have
\begin{equation}
\mu _{2}\equiv \sup_{n}({n^{-1}}\mathrm{Tr}H^{2})=\sup_{n}\int \lambda
^{2}N_{n}(\mathrm{d}\lambda )\leq 4m_{2}\leq 4m_{4}^{1/2}<\infty .
\label{3.22}
\end{equation}
Thus
\[
g_{n}(z)=\int \frac{N_{n}(\mathrm{d}\lambda )}{\lambda -z}=-\frac{1}{z}+%
\widehat{g}_{n}(z),
\]
where
\[
\widehat{g}_{n}(z)=\int \frac{\lambda N_{n}(\mathrm{d}\lambda )}{(\lambda
-z)z}.
\]
In view of (\ref{3.22})
\[
|z\widehat{g}_{n}(z)|\leq |\mathrm{Im\,~}z|^{-1}\int |\lambda |N_{n}(\mathrm{%
d}\lambda )\leq |\mathrm{Im\,~}z|^{-1}m_{4}^{1/4},
\]
i.e. the asymptotic relation
\begin{equation}
g_{n}^{-1}(z)=-z\left( 1+O\left( \frac{1}{|\mathrm{Im\,~}z|}\right) \right)
,\ \mathrm{Im\,~}z\rightarrow \infty  \label{3.21.1}
\end{equation}
holds uniformly in $n$. We have also the simple bound
\begin{equation}
|g_{n}(z)|\leq |\mathrm{Im}z|^{-1}  \label{g_n}
\end{equation}
following from (\ref{3.5}) and (\ref{3.8}) and, in addition, according to
Proposition \ref{pr:3.1} and (\ref{3.17}), the bounds
\begin{equation}
|\delta _{2,n}(z)|\leq m_{4}^{1/4}|\mathrm{Im}z|^{-1},  \label{delta_n}
\end{equation}
\begin{equation}
z\delta _{2,n}(z)=n^{-1}\mathrm{Tr}H_{2}zG(z)=n^{-1}\mathrm{Tr}%
H_{2}(-1+HG(z)).  \label{lta}
\end{equation}
Hence, in view of (\ref{3.22})
\begin{equation}
\begin{array}{c}|z\delta _{2,n}(z)|\leq (n^{-1}\mathrm{Tr}H_{2}^{2})^{1/2}+(n^{-1}\mathrm{Tr}
H_{2}^{2})^{1/2}(n^{-1}\mathrm{Tr}H^{2}G(z)G^{\ast }(z))^{1/2}\leq\\
\leq m_{4}^{1/4}+2m_{4}^{1/2}/y_{0}, \end{array} \label{3.21.2}
\end{equation}
i.e. $z\delta _{2,n}(z)$ is uniformly bounded in $n$.

As a result of above bounds we have for $|\mathrm{Im\,~}z|\geq y_{0}$
uniformly in $n$
\[
||\Delta _{2,n}(z)f_{n}^{-1}(z)G_{1}(z)||=O\left( {\frac{1}{y_{0}}}\right)
,y_{0}\rightarrow \infty
\]
i.e. the matrix $1-\Delta _{2,n}(z)f_{n}^{-1}(z)G_{1}(z)$ is invertible
uniformly in $n$ and there is $y_{0}$ independent of $n$ and such that for $|%
\mathrm{Im\,~}z|\geq y_{0}$
\begin{equation}
||(1+\Delta _{2,n}(z)f_{n}^{-1}(z)G_{1}(z))^{-1}||\leq 2.  \label{3.21.3}
\end{equation}
Thus (\ref{3.21}) is equivalent to
\[
\begin{array}{c}\langle G(z)\rangle =(1+\Delta
_{2,n}(z)f_{n}^{-1}(z)G_{1}(z))^{-1}G_{1}(z)+\\ (1+\Delta
_{2,n}(z)f_{n}^{-1}(z)G_{1}(z))^{-1}f_{n}^{-1}(z)R_{1,n}(z)\end{array}
\]
or to
\[
\langle G(z)\rangle =G_{1}\left( z-\Delta _{2,n}(z)f_{n}^{-1}(z)\right)
+(1+\Delta _{2,n}(z)f_{n}^{-1}(z)G_{1}(z))^{-1}f_{n}^{-1}(z)R_{1,n}(z).
\]
Applying to this relation the operation $n^{-1}\mathrm{Tr}$ we obtain
\begin{equation}
f_{n}(z)=f_{1,n}(z-\Delta _{2,n}(z)f_{n}^{-1}(z))+r_{1,n}(z),  \label{3.23}
\end{equation}
where
\begin{equation}
f_{1,n}(z)=n^{-1}\mathrm{Tr}G_{1}(z)=\int \frac{N_{1,n}(\mathrm{d}\lambda )}{%
\lambda -z}  \label{3.24}
\end{equation}
is the Stieltjes transform of the normalized counting measure of $H_{1,n}$
in (\ref{2.1}) and
\begin{equation}
r_{1,n}(z)=n^{-1}\mathrm{Tr}(1+\Delta
_{2,n}(z)f_{n}^{-1}(z)G_{1}(z))^{-1}f_{n}^{-1}(z)R_{1,n}(z),  \label{3.241}
\end{equation}
where $R_{1,n}(z)$ is defined in (\ref{R1n}). We show in the next Theorem
\ref{th:3.2}
that there exists a sufficiently big $y_{0}>0$ and $C(y_{0})>0$, both
independent of $n$ and such that if $z\in E(y_{0}),$ where $E(y_{0})$ is
defined in (\ref{E}), then the variances
\begin{equation}
v_{1}(z)=\langle |g_{n}^{\circ }(z)|^{2}\rangle ,\ v_{2}(z)=\langle |\delta
_{2,n}^{\circ }(z)|^{2}\rangle  \label{3.25}
\end{equation}
admit the bounds
\begin{equation}
v_{1}(z)\leq \frac{C(y_{0})}{n^{2}},\ v_{2}(z)\leq \frac{C(y_{0})}{n^{2}}.
\label{3.26}
\end{equation}
These bounds, Proposition \ref{pr:3.1},
(\ref{3.21.3}), and Schwartz inequality for
the expectation $\langle ...\rangle $ imply that uniformly in $n$ and in $%
z\in E(y_{0})$%
\[
|r_{1,n}(z)|\leq \frac{2C^{1/2}(y_{0})}{n}(1+y_{0}^{-1})\langle
|f_{n}^{-2}(z)n^{-1}\mathrm{Tr}G(z)G^{\ast }(z)|^{2}\rangle ^{1/2}.
\]
In view of (\ref{3.20}), (\ref{3.21.1}) and the identity $zG(z)=-1+HG(z)$
we have
\[
f_{n}^{-1}(z)G(z)=-z(1+O(y_{0}^{-1}))G(z)=(1+O(y_{0}^{-1}))(1-HG(z)),
\]
and since, by (\ref{3.4}), (\ref{3.5}) and (\ref{TrH2})
\[ \begin{array}{c}
|\langle n^{-1}\mathrm{Tr}HG(z)\rangle |\leq y_{0}^{-1}\langle n^{-1}\mathrm{%
Tr}H^{2}\rangle \leq\\ \leq 2m_{4}^{1/4}y_{0}^{-1},\;|\langle n^{-1}
\mathrm{Tr}H^{2}G(z)G^{\ast }(z)\rangle |\leq 4m_{4}^{1/2}y_{0}^{-2},\end{array}
\]
we obtain that for $z\in E(y_{0})$%
\begin{equation}
|r_{1,n}(z)|\leq \frac{C_{1}(y_{0})}{n},  \label{3.261}
\end{equation}
where $C_{1}(y_{0})$ is independent of $n$ and is bounded in $y_{0}$.

Furthemore, the bounds (\ref{g_n}) and (\ref{delta_n}) imply that sequences $%
\{f_{n}(z)\}$ and $\{\Delta _{2,n}(z)\}$ are analytic and uniformly in $n$
bounded for $|\mathrm{Im\,~}z|\geq y_{0}>0$. Thus the sequences are compact
with respect to uniform convergence on compacts of the domain
\begin{equation}
D(y_{0})=\{z\in \mathbb{C}:|\mathrm{Im\,~}z|\geq y_{0}>0\}.  \label{3.28}
\end{equation}
In addition, according to the hypothesis of the theorem, the normalized
counting measures $N_{1,n}$ of matrices $H_{1,n}$ converge weakly to a
limiting probability measure $N_{1}$ Thus, their Stieltjes transforms (\ref
{3.24}) converge uniformly on compacts of (\ref{3.28}) to the Stieltjes
transform $f_{1}$ of $N_{1}$. Hence, if $y_{0}>0$ is large enough, there
exist two analytic in (\ref{3.28}) functions $f$ and $\Delta _{2}$ verifying
the relation
\[
f(z)=f_{1}\left( z-\frac{\Delta _{2}(z)}{f(z)}\right) ,\ |\mathrm{Im\,~}%
z|\geq y_{0}.
\]
This is the first equation of system (\ref{2.4}). The second equation of the
system follows from the argument above in which the roles $H_{1}$ and $H_{2}$
are interchanged, in particular the quantity $\langle n^{-1}\mathrm{Tr}%
H_{1}G(z)\rangle $ is denoted $\Delta _{1,n}(z)$. As for the third equation,
it is just the limiting form of the identity
\begin{equation}
\langle n^{-1}\mathrm{Tr}(H_{1,n}+H_{2,n}-z)G(z)\rangle =1.  \label{id}
\end{equation}
Thus, we have derived system (\ref{2.4}). Its unique solubility in domain (%
\ref{E}) where $y_{0}$ is large enough is proved in Proposition \ref{pr:3.3}.
Besides, all three functions $f_{n}$, $\Delta _{r,n},r=1,2$ defined in (\ref
{3.20}) are a priory analytic for $|\mathrm{Im\,~}z|>0$. Thus, their limits $%
f,\Delta _{r},r=1,2$ are also analytic for non-real $z$. In view of the weak
compactness of probability measures and the continuity of the one-to-one
correspondence between nonnegative measures and their Stieltjes transforms
(see Proposition \ref{pr:2.1}(v))
there exists a unique nonnegative measure $N$ such
that $f$ admit the representation (\ref{st}). The measure $N$ is a
probability measure in view of (\ref{3.21.1}) and.(\ref{as}).

We conclude that the whole sequence $\{f_{n}\}$ of expectations (\ref{3.20})
of the Stieltjes transforms $g_{n}$ (\ref{3.15}) of measures (\ref{NCM})
converges uniformly on compacts of $D(y_{0}),$ where $D(y_{0})$ is defined
in (\ref{3.28}), to the limiting function $f$ verifying (\ref{2.4}). This
result, Theorem \ref{th:3.2}
and the Borel-Cantelli lemma imply that the sequence $%
\{g_{n}(z)\}$ converges with probability 1 to $f(z)$ for any fixed $z\in
D(y_{0}).$ Since the convergence of a sequence of analytic functions on any
countable set having an accumulation point in their common domain of
definition implies the uniform convergence of the sequence on any compact of
the domain, we obtain the convergence $g_{n}$ to $f$ with probability 1 on
any compact of $D(y_{0})$. Due to the continuity of the one-to-one
correspondence between probability measures and their Stieltjes transforms
(see Proposition \ref{pr:2.1}(v))
the normalized eigenvalue counting measure (\ref
{NCM}) of the eigenvalues of random matrix (\ref{2.1}) converge weakly with
probability 1 to the nonrandom measure $N$ whose Stieltjes transform (\ref
{st}) satisfies (\ref{2.4}).$\blacksquare $

\begin{theorem}
\label{th:3.2} Let $H_{n}$ be the random matrix of
the form (\ref{2.1}) satisfying the condition of Theorem \ref{th:3.1}.
Denote
\begin{equation}
g_{n}(z)=n^{-1}\mathrm{Tr}(H_{n}-z)^{-1},\ \delta _{r,n}(z)=n^{-1}\mathrm{Tr}%
H_{r,n}(H_{n}-z)^{-1},r=1,2.  \label{3.29}
\end{equation}
Then there exist $y_{0}$ and $C(y_{0})$, both positive and independent of $n$
and such that the variances of random variables (\ref{3.29}) admit the
bounds for $|\mathrm{Im\,~}z|\geq y_{0}$
\begin{equation}
\langle |g_{n}(z)-\langle g_{n}(z)\rangle |^{2}\rangle \leq \frac{C(y_{0})}{%
n^{2}}  \label{3.30}
\end{equation}
\begin{equation}
\langle |\delta _{r,n}(z)-\langle \delta _{r,n}(z)\rangle |^{2}\rangle \leq
\frac{C(y_{0})}{n^{2}},r=1,2,.  \label{3.31}
\end{equation}
if $z\in E(y_{0})$, where $E(y_{0})$ is defined in (\ref{E}).
\end{theorem}

\textit{Proof.} Because of the symmetry of the problem with respect to $%
H_{1} $ and $H_{2}$ in (\ref{2.1}) it suffices to prove (\ref{3.31}) for,
say, $\delta _{2,n}(z)$. Besides, we will use bellow the notations $g(z)$
and $\delta (z)$ for $g_{n}(z)$ and $\delta _{2,n}(z)$ and the notations 1
and 2 for two values $z_{1}$ and $z_{2}$ of the complex spectral parameter $%
z $. We assume that $|\mathrm{Im\,~}z_{1,2}|\geq y_{0}>0$.

We will use the same approach as in the proof of Theorem \ref{th:3.1},
i.e. we will
derive and study certain relations obtained by using Proposition \ref{pr:3.2}
and the resolvent identity.

Consider the matrix
\begin{equation}
V_{1}=\langle g^{\circ }(1)G(2)\rangle ,  \label{3.31.1}
\end{equation}
where $g^{\circ }(1)=g(1)-\langle g(1)\rangle $. Its clear that $n^{-1}%
\mathrm{Tr}V_{1}$ for $z_{1}=z$ and $z_{2}=\overline{z}$ is the variance (%
\ref{3.30}), that we denoted $v_{1}(z)$ in (\ref{3.25}):
\begin{equation}
\langle |g^{\circ }(z)|^{2}\rangle =n^{-1}\mathrm{Tr}V_{1}|_{z_{1}=z,z_{2}=%
\overline{z}}=v_{1}(z).  \label{3.32}
\end{equation}
In view of the resolvent identity (\ref{3.14}) for the pair $(H_{1},H)$ we
have
\begin{equation}
V_{1}=-G_{1}(2)W,  \label{3.33}
\end{equation}
\begin{equation}
W=\langle g^{\circ }(1)H_{2}G(2)\rangle .  \label{3.34}
\end{equation}
Applying Proposition \ref{pr:3.2} to the function
\[
\Phi (M)=G_{aa}^{\circ }(1)(MG(2))_{cd},
\]
where $G(z)=(H_{1}+M-z)^{-1}$, and
\begin{eqnarray*}
G^{\circ }(z) &=&G(z)-\langle G(z)\rangle = \\
&&(H_{1}+M-z)^{-1}-\int_{U(n)}(H_{1}+U^{\ast }BU-z)^{-1}dU,
\end{eqnarray*}
we obtain the relation
\[
\begin{array}{c}
-\langle (G(1)[X,H_{2}]G(1))_{aa}(H_{2}G(2))_{cd}\rangle +\langle
G_{aa}^{\circ }(1)([X,H_{2}]G(2))_{cd}\rangle - \\
-\langle G_{aa}^{\circ }(1)(H_{2}G(2)[X,H_{2}]G(2))_{cd}\rangle =0,
\end{array}
\]
where the operation $[...,...]$ is defined in (\ref{com}). Choosing as $X$
the Hermitian matrix having only the $(c,j)$-th and $(j,c)$ non-zero
entries, we obtain from the above relation the following one:
\[
-\langle G_{ac}(1)(H_{2}G(1))_{ja}(H_{2}G(2))_{cd}\rangle +\langle
(G(1)H_{2})_{ac}G_{ja}(1)(H_{2}G(2))_{cd}\rangle+\]
\[
+\langle G_{aa}^{\circ}(1)\delta _{cc}(H_{2}G(2))_{jd}\rangle
-\langle G_{aa}^{\circ }(1)(H_{2})_{cc}G_{jd}(2)\rangle-\]
\[-\langle G_{aa}^{\circ }(1)(H_{2}G(2))_{cc}(H_{2}G(2))_{jd}\rangle
+\langle G_{aa}^{\circ }(1)(H_{2}G(2)H_{2})_{cc}G_{jd}(2)\rangle =0
\]
Applying to this relation the operation $n^{-1}\sum\limits_{ac}$ and taking
into account that
\[
g^{\circ }=n^{-1}\sum\limits_{a}G_{aa}^{\circ },
\]
we have
\begin{equation} \begin{array}{c}
n^{-2}\langle \lbrack G^{2}(1),H_{2}]H_{2}G(2)\rangle +\langle g^{\circ}(1)
H_{2}G(2)\rangle +\\ +\langle g^{\circ }(1)k(2)G(2)\rangle -\langle g^{\circ
}(1)\delta (2)H_{2}G(2)\rangle =0,\end{array}  \label{3.35}
\end{equation}
where
\begin{equation}
k(z)=n^{-1}\mathrm{Tr}K(z),K(z)=BG_{U}(z)B-B,\ \ G_{U}(z)=UG(z)U^{\ast }.
\label{3.37}
\end{equation}
Introducing the centralized quantity (cf.(\ref{3.19}))
\begin{equation}
k^{\circ }=k-\langle k\rangle ,  \label{3.38}
\end{equation}
and using our notations (\ref{3.17}) and (\ref{3.20}), we can rewrite (\ref
{3.35}) as
\begin{equation}
(1-\Delta (2))W=-\langle k(2)\rangle V_{1}+R,  \label{3.39}
\end{equation}
where
\begin{equation}
R=\langle g^{\circ }(1)\delta ^{\circ }(2)H_{2}G(2)\rangle -\langle g^{\circ
}(1)k^{\circ }(2)G(2)\rangle -T_{1},  \label{3.40}
\end{equation}
and
\[
T_{1}=n^{-2}\langle \lbrack G^{2}(1),H_{2}]H_{2}G(2)\rangle .
\]
In view of the uniform in $n$ bound (\ref{3.21.2})), the function $1-\Delta
(z)$ is uniformly in $n$ bounded away from zero. Thus we have from (\ref
{3.33}), (\ref{3.34}) and (\ref{3.39})
\begin{equation}
V_{1}=\left( 1-\langle k(2)\rangle (1-\Delta (2))^{-1}G_{1}(2)\right)
^{-1}(1-\Delta (2))^{-1}G_{1}(2)R.  \label{3.41}
\end{equation}
According to (\ref{3.37}), (\ref{3.7}) and (\ref{3.1}), we have uniformly in
$n$
\begin{equation}
|k(z)|\leq y_{0}^{-1}n^{-1}\mathrm{\ Tr}B^{2}+|n^{-1}\mathrm{Tr}B|\leq
y_{0}^{-1}m_{4}^{1/2}+m_{4}^{1/4}<\infty .  \label{k_n}
\end{equation}
This bound and universal bound (\ref{3.7}) imply that the matrix $(1-\langle
k(z)\rangle (1-\Delta (z))^{-1}G_{1}(z))$ is uniformly in $n$ invertible if $%
|\mathrm{Im\,~}z|\geq y_{0}$ and $y_{0}$ is large enough, and hence the
matrix
\[
Q=\left( 1-\langle k(z)\rangle (1-\Delta (z))^{-1}G_{1}(z)\right)
^{-1}(1-\Delta (z))^{-1}G_{1}(z)
\]
admits the following bound for $|\mathrm{Im\,~}z|\geq y_{0}$ and
sufficiently large $y_{0}$
\begin{equation}
||Q||\leq \frac{C}{y_{0}},  \label{3.42}
\end{equation}
where $C$ is an absolute constant.

Setting now in (\ref{3.41}) $z_{1}=z$, $z_{2}=\overline{z}$ and applying to
this relation the operation $n^{-1}\mathrm{Tr}$ we obtain in the l.h.s. the
variance $v_{1}(z)$ because of (\ref{3.32}). As for the r.h.s., its terms
can be estimated as follows in view of (\ref{3.40}):

\begin{enumerate}
\item[(i)]
\begin{equation}
|\langle g^{\circ }(1)\delta ^{\circ }(2)n^{-1}\mathrm{Tr}QH_{2}G(2)\rangle
|\leq \alpha _{12}(y_{0})v_{1}^{1/2}v_{2}^{1/2},  \label{3.43}
\end{equation}
where $v_{2}$ is defined in (\ref{3.25}) and because, according to (\ref{3.1}%
), (\ref{3.4}), (\ref{3.7}) and (\ref{3.42}),
\begin{equation} \begin{array}{c}
|n^{-1}\mathrm{Tr}QH_{2}G(2)|\leq (n^{-1}\mathrm{Tr}Q^{\ast }Q)^{1/2}(n^{-1}%
\mathrm{Tr}H_{2}^{2}G(2)G^{\ast}(2))^{1/2}\leq\\ \leq Cy_{0}^{-2}m_{4}^{1/4}
\equiv\alpha _{12}(y); \end{array}  \label{al12}
\end{equation}

\item[(ii)]
\begin{equation}
|\langle g^{\circ }(1)k^{\circ }(2)n^{-1}\mathrm{Tr}QG(2)\rangle |\leq
\alpha _{13}(y_{0})v_{1}^{1/2}v_{3}^{1/2},  \label{3.44}
\end{equation}
where
\begin{equation}
v_{3}=\langle |k^{\circ }(z)|^{2}\rangle  \label{v_3}
\end{equation}
because
\begin{equation} \begin{array}{c}
|n^{-1}\mathrm{Tr}QG(2)|\leq (n^{-1}\mathrm{Tr}Q^{\ast }Q)^{1/2}(n^{-1}%
\mathrm{Tr}G(2)G^{\ast }(2))^{1/2}\leq\\ \leq Cy_{0}^{-2}\equiv
\alpha _{13}(y_{0}); \end{array}\label{al13}
\end{equation}

\item[(iii)]
\[
|n^{-3}\mathrm{Tr}(Q[G^{2}(1),H_{2}]H_{2}G(2))|\leq
Cm_{4}^{1/2}y_{0}^{-4}n^{-2}\equiv \frac{\beta _{1}(y_{0})}{n^{2}}.
\]
\end{enumerate}

Thus we obtain the inequality
\begin{equation}
v_{1}\leq \alpha _{12}(y_{0})v_{1}^{1/2}v_{2}^{1/2}+\alpha
_{13}(y_{0})v_{1}^{1/2}v_{3}^{1/2}+\frac{\beta _{1}(y_{0})}{n^{2}},
\end{equation}
where $\alpha _{12}$, $\alpha _{13}$ and $\beta _{1}$ are independent on $n$
and vanish as $y_{0}\rightarrow \infty $.

Now we are going to derive analogous inequalities for $v_{2}$ and $v_{3}$
defined in (\ref{3.25}) and in (\ref{v_3}) and to obtain the system
\begin{equation}
v_{i}\leq \sum_{j=1,j\neq i}^{3}\alpha _{ij}v_{i}^{1/2}v_{j}^{1/2}+\frac{%
\beta _{i}(y_{0})}{n^{2}},\ i=1,2,3.  \label{3.46}
\end{equation}
To get the second inequality of the system we consider the matrix (cf. (\ref
{3.31.1}))
\begin{equation}
V_{2}=\langle \delta ^{\circ }(1)H_{2}G(2)\rangle .  \label{3.47}
\end{equation}
Applying to $V_{2}$ operation $n^{-1}\mathrm{Tr}$ and setting $z_{1}=z$, $%
z_{2}=\overline{z,}$ we obtain the variance $v_{2}$ of (\ref{3.26}). On the
other hand, using Proposition \ref{pr:3.2} for the function
\[
\Phi (M)=(MG(1))_{aa}^{\circ }(MG(2))_{cd},
\]
we obtain, after performing in essence the same procedure as that used in
the derivation of (\ref{3.35}), in particular, choosing the Hermitian matrix
$X$ with only the $(c,j)$-th and $(j,c)$ non-zero entries,
\begin{equation}
v_{2}=-\langle g(2)\delta ^{\circ }(1)k(2)\rangle +\langle \delta ^{\circ
}(1)\delta ^{2}(2)\rangle -T_{2},  \label{3.47.1}
\end{equation}
where
\begin{equation}
T_{2}=\langle n^{-3}\mathrm{Tr}([G_{U}(1),K(1)]BG(2))\rangle  \label{T_2}
\end{equation}
and $K(z)$, $k(z)$ are defined in (\ref{3.37}). Using again centralized
quantities (\ref{3.19}) and (\ref{3.38}), we can write
\[
\langle g(2)\delta ^{\circ }(1)k(2)\rangle =\langle g^{\circ }(2)\delta
^{\circ }(1)k(2)\rangle +\langle g(2)\rangle \langle \delta ^{\circ
}(1)k^{\circ }(2)\rangle
\]
and
\[
\langle \delta ^{\circ }(1)\delta ^{2}(2)\rangle =\langle \delta ^{\circ
}(1)\delta ^{\circ }(2)\delta (2)\rangle +\langle \delta ^{\circ }(1)\delta
^{\circ }(2)\rangle \langle \delta (2)\rangle .
\]
Thus, in view of (\ref{g_n}), (\ref{delta_n}), (\ref{k_n}), and Schwarz
inequality we have the bounds
\[
\langle g(2)\delta ^{\circ }(1)k(2)\rangle \leq
v_{1}^{1/2}v_{2}^{1/2}m_{4}^{1/4}(1+m_{4}^{1/4}y_{0}^{-1})+v_{2}^{1/2}v_{3}^{1/2}y_{0}^{-1},
\]
and
\[
\langle \delta ^{\circ }(1)\delta ^{2}(2)\rangle \leq
2v_{2}m_{4}^{1/4}y_{0}^{-1}.
\]
These bounds and analogously obtained bound for $T_{2}$ in (\ref{T_2}) lead
for $m_{4}^{1/4}y_{0}^{-1}\leq 1/4$ to the second inequality (\ref{3.46}),
in which
\begin{equation}
\alpha _{21}(y_{0})=4m_{4}^{1/4},\ \alpha _{23}(y_{0})=2y_{0}^{-1},\ \beta
_{2}=8m_{4}^{1/4}y_{0}^{-2}.  \label{al213}
\end{equation}

To obtain the third inequality of (\ref{3.46}) we may use the same scheme as
above applied to the matrix $V_{3}=\langle k^{\circ }(1)K(2)\rangle $ (cf. (%
\ref{3.31.1}) and (\ref{3.47})). However this requires rather tedious
computations and the existence of the uniformly bounded in $n$ sixth moment $%
m_{6}$ of the measure $N_{2,n}$. For this reason we consider the quantity
\begin{equation}
\langle n^{-1}\mathrm{Tr}(BG_{U}(1)B)^{\circ }G_{U}(2)B\rangle ,
\label{3.48}
\end{equation}
where $G_{U}(z)$ is defined in (\ref{3.37}). As before we would like to
obtain for this quantity a certain relation, basing on the invariance of the
Haar measure with respect to the group shifts. To this end we will introduce
the following function of the unitary matrix $U$:
\[
(BUG(1)U^{\ast }B)_{aa}^{\circ }(UG(2)U^{\ast }B)_{cd},
\]
where $G(z)=(H_{1}+U^{\ast }BU-z)^{-1}$ and we will use the analogue of (\ref
{3.12}) obtained from the left shift invariance of the Haar measure. This
leads to the relation (cf. (\ref{3.35}) and (\ref{3.47.1}))
\begin{equation}
\langle k^{\circ }(1)g(2)K(2)\rangle +\langle k^{\circ }(1)\delta
(2)G_{U}(2)B\rangle -\langle k^{\circ }(1)G_{U}(2)B\rangle -T_{3}=0,
\label{3.49}
\end{equation}
where
\[
T_{3}=n^{-2}\langle G_{U}(1)BK(1)G_{U}(2)B-K(1)BG_{U}(1)G_{U}(2)B\rangle .
\]
We multiply (\ref{3.49}) by $B$ from the left and introduce again the
centralized quantities $g^{\circ }$, $\delta ^{\circ }$ and $k^{\circ }$
defined in (\ref{3.19}) and (\ref{3.38}). We obtain
\[\begin{array}{c}
(1-\Delta (2)-f(2)B)\langle k^{\circ }(1)K(2)\rangle =-\langle k^{\circ
}(1)g^{\circ }(2)BK(2)\rangle+\\ +\langle k^{\circ }(1)\delta ^{\circ
}(2)BG_{U}(2)B\rangle +BT_{3}.  \end{array}
\]
In view of (\ref{3.21.1}) and (\ref{3.21.2}) the imaginary part of the
function $1-\Delta (z)$ is uniformly in $n$ bounded away from zero if $|%
\mathrm{Im\,~}z|$ is large enough. Since $B$ is a Hermitian matrix, the
matrix
\begin{equation}
S=(1-\Delta (2)-f(2)B)^{-1}  \label{3.491}
\end{equation}
admits the bound
\[
||S||=|f(2)|^{-1}\cdot ||((1-\Delta (2))f^{-1}(2)-B)^{-1}||\leq
|f(2)|^{-1}\left| \mathrm{Im}\frac{1-\Delta (2)}{f(2)}\right| ^{-1}.
\]
By using (\ref{3.21}) and (\ref{delta_n}) we find that for $z\in E(y_{0})$,
where $E(y_{0})$ is defined in (\ref{E}) with sufficiently big $y_{0}$, we
have the uniform in $n$ inequality $|f(2)\mathrm{Im}(1-\Delta
(2))f^{-1}(2)|\geq 1/2,$ i.e.
\begin{equation}
||S||\leq 2.  \label{3.50}
\end{equation}
This leads to the relation
\begin{equation}\begin{array}{c}
V_{3}\equiv \langle k^{\circ }(1)K(2)\rangle =-\langle k^{\circ }(1)g^{\circ
}(2)SBK(2)\rangle+\\ +\langle k^{\circ }(1)\delta ^{\circ}(2)
SBG_{U}(2)B\rangle +SBT_{3}.\end{array}  \label{3.51}
\end{equation}
We apply to this relation the operation $n^{-1}\mathrm{Tr}$, set $z_{1}=z$, $%
z_{2}=\overline{z}$ and estimate the contribution of the two first terms of
the r.h.s. as (\ref{3.51}) as above, using in addition (\ref{3.50}). We
obtain
\begin{equation}\begin{array}{l}
|n^{-1}\mathrm{Tr}SBK(2)|\leq 4m_{4}^{1/2}\equiv \alpha
_{31}(y_{0}),\\ |n^{-1}\mathrm{Tr}SBG_{U}(2)B|\leq
4m_{4}^{1/2}y_{0}^{-1}\equiv \alpha _{32}(y_{0}).\end{array}  \label{al312}
\end{equation}
To estimate the third term of the r.h.s. of (\ref{3.51}) we use the identity
\[
SB=-f^{-1}(2)+(1-\Delta (2))f^{-1}(2)S,
\]
the asymptotic relations (\ref{3.21.1}) and (\ref{delta_n}) and the bound (%
\ref{3.50}). This yields the bound $||SB||\leq 4y_{0}$. By using this bound
and the same reasoning as in obtaining other bounds above, we obtain
\[
|n^{-1}\mathrm{Tr}SBT_{3}|\leq \frac{Cm_{4}}{y_{0}^{2}n^{2}}\equiv \frac{%
\beta _{3}}{n^{2}},
\]
where $C$ is an absolute constant.

Let us introduce new variables
\begin{equation}
u_{1}=y_{0}v_{1}^{1/2},\ u_{2}=v_{2}^{1/2},\ u_{3}=v_{3}^{1/2}  \label{uv}
\end{equation}
Then we obtain from (\ref{3.46}) and (\ref{al12}), (\ref{al13}), (\ref{al213}%
), and (\ref{al312}) the system
\begin{equation}
u_{i}^{2}\leq \sum_{j=1,j\neq i}^{3}a_{ij}u_{i}u_{j}+\frac{\gamma _{i}}{n^{2}%
},  \label{3.52}
\end{equation}
in which the coefficients $\{a_{ij},i\neq j\}$ have the form $%
a_{ij}=y_{0}^{-1}b_{ij},$ where $b_{ij}$ are bounded in $y_{0}$ and in $n$
as $y_{0}\rightarrow \infty $ and $n\rightarrow \infty $. By choosing $y_{0}$
sufficiently big (and then fixing it) we can guarantees that $0\leq
a_{ij}\leq 1/4,i\neq j$. Thus summing the three relations (\ref{3.52}) we
can write the result in the form $(\hat{a}u,u)\leq \gamma /n^{2}$ where $%
\gamma =\gamma _{1}+\gamma _{2}+\gamma _{3}$ and ($\hat{a})_{ij}=\delta
_{ij}+(1-\delta _{ij})/4,i,j=1,2,3.$ Since the minimum eigenvalue of the
matrix $\hat{a}$ is $1/2,$ we obtain from (\ref{uv})bounds (\ref{3.30}) and (%
\ref{3.31}).$\blacksquare $

\setcounter{equation}{0}
\setcounter{theorem}{0}
\setcounter{proposition}{0}
\setcounter{lemma}{0}
\setcounter{remark}{0}
\section{Convergence in Probability}

In this Section we prove Theorem \ref{th:main}.
Since, according to Theorem \ref{th:3.2} the
randomness of $U_{n}$ in (\ref{2.1}) (or (\ref{ABU})) provides already
vanishing the variance of the Stieltjes transform of the NCM (\ref{NCM}), we
have only to prove that the additional randomness due to the matrices $A_{n}$
and $B_{n}$ in (\ref{2.1}) does not destroy this property. We will prove
this fact first for $A_{n}$ and $B_{n}$ whose norms are uniformly bounded in
$n$ (see Lemma \ref{l:4.1} below ),
and then we will treat the general case of
Theorem \ref{th:main} by using a certain truncating procedure.

\begin{proposition}
\label{pr:4.1} Let $\{m_{n}\}$ be a sequence of
random non-negative unit measures on the line and $\{s_{n}\}$
be the sequence of their Stieltjes transforms (\ref{St}).
Then the sequence $\{m_{n}\}$ converges weakly in probability to a
nonrandom non-negative unit measure $m$ if and only if the
sequence $\{s_{n}\}$ converges in probability for any fixed $z$
belonging to a compact $K\subset \{z\in C:\mathrm{Im}z>0\}$
to the Stieltjes transform $f$ of the measure $m$.
\end{proposition}
\medskip

\textit{Proof}. Let us prove first the necessity. According to the
hypothesis for any continuous and having a compact support function $\varphi
(\lambda )$ we have
\begin{equation}
\lim_{n\rightarrow \infty }\mathbf{P}\left\{ \left| \int \varphi (\lambda
)m(d\lambda )-\int \varphi (\lambda )m_{n}(d\lambda )\right| >\varepsilon
\right\} \mathbf{=}0.  \label{Pm}
\end{equation}
Let $\chi (\lambda )$ be a continuous function that is equal to 1 if $%
|\lambda |<A$ and is equal to $0$ if $|\lambda |>A+1$ \ for some $A>0.$ Then
\[
|s(z)-s_{n}(z)|\leq \left| \int \frac{\chi (\lambda )m(d\lambda )}{\lambda -z%
}-\int \frac{\chi (\lambda )m_{n}(d\lambda )}{\lambda -z}\right| +\frac{2}{%
\min \{\mathrm{dist}\{z,\pm A\}\}}.
\]

According to (\ref{Pm}) the first term in the r.h.s. of this inequality
converges in probability to zero. Since $A$ is arbitrary, we obtain the
required assertion.

To prove sufficiency we assume that for any $z\in K$

\begin{equation}
\lim_{n\rightarrow \infty }\mathbf{P\{|}s(z)-s_{n}(z)|>\varepsilon \mathbf{%
\}=}0.  \label{Ps}
\end{equation}
This relation and the inequality (cf. (\ref{bound}))
\begin{equation}
|s_{n}(z)|\leq \max_{z\in K}|\mathrm{Im}z|^{-1}\equiv y_{0}^{-1}<\infty
\label{4.1}
\end{equation}
imply that

\begin{equation}
\lim_{n\rightarrow \infty }\mathbf{E\{|}s(z)-s_{n}(z)|\}\mathbf{=}0,
\label{Es}
\end{equation}
i.e. the sequence $\{s_{n}(z)\}$ converges to zero in mean. We have also the
inequality

\begin{equation}
|s_{n}^{^{\prime }}(z)|\leq y_{0}^{-2}<\infty .  \label{4.2}
\end{equation}
Inequalities (\ref{4.1}) and (\ref{4.2}) imply that the sequence $%
\{s_{n}\}_{n=1}^{\infty }$ of random analytic functions is uniformly bounded
and equicontinuous. Thus, for any $\eta >0$ we can construct in $K$ a finite
$\eta $-network, i.e. a set $\{z_{l}\}_{l=1}^{p(\eta )}$ such that for any $%
z\in K$ there exists $z_{l}$ satisfying the inequality $|z-z_{l}|\leq \eta $%
. Then we have for $\phi _{n}(z)\equiv s_{n}(z)-s(z)$, $S_{l}=\{z:|z-z_{l}|%
\leq \eta \}$, and $\eta =y_{0}^{2}\varepsilon /2,$ where $\varepsilon $ is
arbitrary
\[
\sup_{K}|\phi _{n}(z)|=\max_{l=1...p(\eta )}\sup_{z\in K\cap S_{l}}|\phi
_{n}(z)|\leq \varepsilon +\sum_{l=1}^{p(\eta )}|\phi _{n}(z_{l})|,
\]
and hence
\[
\mathbf{E\{}\sup_{K}|\phi _{n}(z)|\}\leq \varepsilon +\sum_{l=1}^{p(\eta )}%
\mathbf{E\{}|\phi _{n}(z_{l})|\}.
\]
This inequality and (\ref{Es}) imply that

\begin{equation}
\lim_{n\rightarrow \infty }\mathbf{E\{}\sup_{z\in K}\mathbf{|}s(z)-s_{n}(z)|%
\mathbf{\}=}0.  \label{Esup}
\end{equation}

Assume now that the statement is false, i.e. the sequence $\{m_{n}\}$ does
not converges weakly in probability to $m$. It means that there exists a
continuous function $\varphi $ of a compact support, a subsequence $%
\{n_{k}\} $ and some $\varepsilon >0$ such that
\begin{equation}
\lim_{n_{k}\rightarrow \infty }\mathbf{P}\left\{ \left| \int \varphi
(\lambda )m(d\lambda )-\int \varphi (\lambda )m_{n_{k}}(d\lambda )\right|
\geq \varepsilon \right\} \mathbf{=}\xi >0.  \label{4.2_1}
\end{equation}
On the other hand, we have from (\ref{Esup}) and the Tchebyshev inequality
that for any $r$ there exists an integer $n(r)$ such that for $n\geq n(r)$
\begin{equation}
\mathbf{P}\left\{ \sup_{z\in K}|\phi _{n}(z)|\leq r^{-1}\right\} \geq 1-\xi
/2.  \label{4.2_2}
\end{equation}
Hence, one can select from the sequence $\{n_{k}\}$ a subsequence $%
\{n_{k^{\prime }}\}$ such that inequalities (\ref{4.2_1}) and (\ref{4.2_2})
are both satisfied. Denote by $\mathcal{A}$ and by $\mathcal{B}$ the events
whose probabilities are written in the l.h.s. of (\ref{4.2_1}) and (\ref
{4.2_2}). Then $\mathbf{P}\{\mathcal{A}\cap \mathcal{B}\}\geq $ $\mathbf{P}\{%
\mathcal{A}\}+$ $\mathbf{P}\{\mathcal{B}\}-1\geq \xi /2$. Hence, for any $%
n_{k^{\prime }}$ there exists a realization $\omega _{n_{k}^{\prime }}$
belonging to the both sets $\mathcal{A}$ and $\mathcal{B}$, i.e. for which
the both inequalities
\begin{equation}
\left| \int \varphi (\lambda )m(d\lambda )-\int \varphi (\lambda
)m_{n_{k}^{\prime }}(d\lambda )\right| \geq \varepsilon ,\sup_{z\in K}|\phi
_{n_{k}^{\prime }}(z)|\leq r^{-1}  \label{4.2_3}
\end{equation}
are valid. In view of the compactness of the family of the random analytic
functions $\{s_{n}\}$ with respect to the uniform in $K$ convergence and the
weak compactness of the family of random measure $\{m_{n}\}$ there exists a
subsequence $\{n_{k}^{^{\prime \prime }}\}$ of $\{n_{k}^{^{\prime }}\}$ and
a subsequence of realizations $\{\omega _{n_{k}^{\prime \prime }}\}$ such
that the subsequence $\{m_{n_{k}^{\prime \prime }}\}$ corresponding to these
realizations converges weakly to a certain measure $\widetilde{m}$ and we
have in view of (\ref{4.2_1})
\begin{equation}
\left| \int \varphi (\lambda )m(d\lambda )-\int \varphi (\lambda )\widetilde{%
m}(d\lambda )\right| \geq \varepsilon >0.  \label{4.2_4}
\end{equation}
On the other hand, in view of (\ref{4.2_3}) and the continuity of the
correspondence between measures and their Stieltjes transforms (see
Proposition \ref{pr:2.1}(v)),
the subsequence $\{s_{n_{k}^{\prime \prime }}\}$
converges uniformly on $K$ to $s(z)$, the Stieltjes transform of the measure
$m$. This is incompatible with (\ref{4.2_4}), because of the one-to-one
correspondence between measures and their Stieltjes transforms.
$\blacksquare$

\begin{remark}
\label{prem:1}{\rm Since the Stieltjes transforms of non-negative and
normalized to unity measures are analytic and bounded for non-real $z$, we
can replace the requirement of their convergence for any $z$ belonging to a
certain compact of $\mathbb{C}_{\pm }$ by the convergence for any $z$
belonging to any interval of the imaginary axis, i.e. for $z=iy,$ $y\in
\lbrack y_{1},y_{2}],$ $y_{1}>0.$
}\end{remark}

\begin{remark}
\label{prem:2}
{\rm The arguments, used in the proof of the proposition prove also that if $%
\{m_{n}\}$ is a sequence of random non-negative measures converging weakly
in probability to a nonrandom non-negative measure $m$, then the Stieltjes
transforms $s_{n}$ of $m_{n}$ and the Stieltjes transform $s$ of $m$ are
related as follows
\begin{equation}
\lim_{n\rightarrow \infty }\mathbf{E}\{\sup_{z\in K}|s_{n}(z)-s(z)|\}=0
\label{4.2_5}
\end{equation}
for any compact $K$ of $\mathbb{C}_{\pm }$.
}\end{remark}

\begin{lemma}
\label{l:4.1} Let $H_{n}$ be the random $n\times n$
matrix of the form (\ref{2.1}) in which $A_{n}$ and $B_{n}$ are random
Hermitian matrices, $U_{n}$ and $V_{n}$ are random unitary matrices
distributed each according to the normalized to unity Haar measure on $%
\mathbf{U}(n)$ and $A_{n}$, $B_{n}$, $U_{n}$ and $V_{n}$ are mutually
independent. Assume that the normalized counting measures $N_{r,n},r=1,2$ of
matrices $A_{n}$ and $B_{n}$ converge in probability as $n\rightarrow \infty
$ to non-random non-negative unit measures $N_{r},r=1,2$ respectively and
that
\begin{equation}
\sup_{n}||A_{n}||\leq T<\infty ,\ \sup_{n}||B_{n}||\leq T<\infty .
\label{4.3}
\end{equation}
Then the normalized counting measure of $H_{n}$ converges in probability to
a non-random unity measure $N$ whose Stieltjes transform $f(z)$ is a unique
solution of system (\ref{2.4}) in the class of functions $f(z)$, $\Delta
_{r}(z),r=1,2$ analytic for $\mathrm{Im\,~}z\neq 0$ and satisfying
conditions (\ref{bound}) - (\ref{as}) and (\ref{2.6})
\end{lemma}
\medskip

\textit{Proof.} In view of Proposition \ref{pr:4.1}
it suffices to show that $%
\lim_{n\rightarrow \infty }\mathbf{E}\{|g_{n}(z)-f(z)|\}=0$ for any $z$
belonging to a certain compact of $\mathbb{C}_{\pm }$. Moreover, according
to Remark \ref{prem:1} after Proposition \ref{pr:4.1},
we can restrict ourselves to a certain
interval of the imaginary axis, i.e. to
\begin{equation}
z=iy,y\in \lbrack y_{1},y_{2}],0<y_{1}<y_{2}<\infty .  \label{z}
\end{equation}
Since the condition (\ref{4.3}) of the lemma implies evidently the condition
(\ref{3.1}) of Theorem \ref{th:3.1} and Theorem \ref{th:3.2},
all the results obtained in
these theorems are valid in our case for any fixed realization of random
matrices $A_{n}$ and $B_{n}$. In addition, all $n$-independent estimating
quantities entering various bounds in the proofs of these theorems and
depending on the forth moment $m_{4}$ in (\ref{3.1}) and on $y_{0}$ will
depend now on $T$ and on $y_{1}$ and $y_{2}$ in (\ref{z}), but not on
particular realizations of random matrices $A_{n}$ and $B_{n}.$ We will
denote below all these quantities simply by the unique symbol $C$ that may
have different value in different formulas.

In particular, denoting as above by $\langle ...\rangle $ the expectation
with respect to the Haar measure and using (\ref{3.26}), we can write that
\[
\mathbf{E}\{|g_{n}(z)-\langle g_{n}(z)\rangle |\}\leq \mathbf{E}%
\{|v_{1}^{1/2}(z)|\}\leq \frac{C}{n}.
\]
Thus, it suffices to show that
\begin{equation}
\lim_{n\rightarrow \infty }\mathbf{E}\{|\langle g_{n}(z)\rangle
-f(z)|\}=0,z=iy,\ y\in \lbrack y_{1},y_{2}],  \label{Exp}
\end{equation}
where $y_{1}$ is big enough. Introduce the quantities
\begin{equation}
\gamma _{n}(y)=iy(\langle g_{n}(iy)\rangle -f(iy)),\ \gamma
_{r,n}(y)=\langle \delta _{r,n}(iy)\rangle -\Delta _{r}(iy),\ r=1,2.
\label{gam_r}
\end{equation}
By using the second equation of system (\ref{2.4}) we can write the identity
\begin{equation}
\gamma _{n}(y)=iy[f_{2}(iy-t_{1,n}(y))-f_{2}(iy-t_{1}(y))]+\varepsilon
_{1,n}(y),  \label{Exp1}
\end{equation}
where
\begin{equation}
\varepsilon _{1,n}(y)=iy[\langle g_{n}(iy)\rangle -f_{2}(iy-t_{1,n}(y))],
\label{eps_1}
\end{equation}
\begin{equation}
t_{1,n}(y)=\frac{\langle \delta _{1,n}(iy)\rangle }{\langle g_{n}(iy)\rangle
},\ t_{1}(y)=\frac{\Delta _{1}(iy)}{f(iy)}.  \label{t}
\end{equation}
We have
\begin{equation} \begin{array}{c}
\mathbf{E}\{|\varepsilon _{1,n}(y)|\}\leq y_{2}\mathbf{E}\{|\langle
g_{n}(iy)\rangle -g_{2,n}(iy-t_{1,n}(y))|\}+\\ \mathbf{E}%
\{|g_{2,n}(iy-t_{1,n}(y))-f_{2}(iy-t_{1,n}(y))|\}.\end{array} \label{4.31}
\end{equation}
The analogues of (\ref{3.23}) - (\ref{3.24}) in our case are:
\begin{equation}
\langle g_{n}(z)\rangle =g_{2,n}(z-\langle \delta _{1,n}(z)\rangle \langle
g_{n}(z)\rangle ^{-1})+\widehat{r}_{1,n}(z),  \label{4.17}
\end{equation}
where
\[
g_{2,n}(z)=n^{-1}\mathrm{Tr}G_{2}(z)=\int \frac{N_{2,n}(\mathrm{d}\lambda )}{%
\lambda -z},
\]
is the Stieltjes transform of random NCM $N_{2,n}$ of $H_{2,n}$,
\[ \begin{array}{c}
\widehat{r}_{1,n}(z)=-\langle g_{n}^{\circ }(z)n^{-1}\mathrm{Tr}%
P^{-1}\langle g_{n}(z)\rangle ^{-1}G(z)\rangle-\\ -\langle
\delta _{1,n}^{\circ}(z)n^{-1}\mathrm{Tr}P^{-1}\langle g_{n}(z)
\rangle ^{-1}G_{2}(z)G(z)\rangle,\end{array}
\]
the symbol $\langle ...\rangle $ denotes the expectation with respect the
Haar measure on $U(n),$ $P=1-G_{2}(z)t_{1,n}(z),$ and
\begin{equation}
g_{n}^{\circ }(z)=g_{n}(z)-\langle g_{n}(z)\rangle ,\ \delta _{1,n}^{\circ
}(z)=\delta _{1,n}(z)-\langle \delta _{1,n}(z)\rangle  \label{4.18}
\end{equation}
are the respective random variables centralized by the partial expectations
with respect to the Haar measure. In addition, we have the analogue of
 (\ref{3.261})
\[
\left| \widehat{r}_{1,n}(z)\right| \leq \frac{C}{n}.
\]
This leads to the following bound for the first term in the r.h.s. of (\ref
{4.31}):
\[
\mathbf{E}\{|\langle g_{n}(iy)\rangle -g_{1,n}(iy-t_{2,n}(y)|\}\leq \mathbf{E%
}\{|\widehat{r}_{1,n}(iy)|\}\leq \frac{C}{n}.
\]
To show that the second term also vanishes as $n\rightarrow \infty $, we use
the analogues of (\ref{3.21.1}) and (\ref{3.21.2})
\[
\left| \langle g_{1,n}(iy)\rangle +\frac{1}{iy}\right| \leq \frac{T}{y^{2}}%
,\ |\delta _{2,n}(iy)|\leq \frac{T}{y},
\]
which imply that
\begin{equation}
|t_{1,n}(y)|\leq 2T,  \label{t_1}
\end{equation}
if $y_{1}$ is big enough. Thus
\[
\mathbf{E}\{|g_{2,n}(iy-t_{1,n}(y))-f_{2}(iy-t_{1,n}(y))|\}\leq \sup_{|\zeta
|\leq T}\mathbf{E}\{|g_{2,n}(iy+\zeta )-f_{1}(iy+\zeta )|\}.
\]
The r.h.s of this inequality tends to zero as $n\rightarrow \infty $ in view
of the hypothesis of Theorem \ref{th:main}
and Remark \ref{prem:2} after Proposition \ref{pr:4.1}. Thus,
there exist $0<y_{1}<y_{2}<\infty $ such that for all $y\in \lbrack
y_{1},y_{2}],$ $\lim_{n\rightarrow \infty }\mathbf{E}\{|\varepsilon
_{1,n}(y)|\}=0.$ Analogous arguments show that $\lim_{n\rightarrow \infty }%
\mathbf{E}\{|\varepsilon _{2,n}(y)|\}=0$, where $\varepsilon _{2,n}(y)$ is
defined in (\ref{eps_1}) and in (\ref{t}) where the indices 1 and 2 are
interchanged. Thus we have
\begin{equation}
\lim_{n\rightarrow \infty }\mathbf{E}\{|\varepsilon _{r,n}(y)|\}=0,\,\,r=1,2.
\label{eps_r}
\end{equation}
Consider now the first term in the l.h.s. of (\ref{Exp1}). In view of (\ref
{f12}) we can write this term in the form
\begin{equation}
\lbrack f_{2}(iy-t_{1,n}(y))-f_{2}(iy-t_{1}(y))]=-\frac{\langle \delta
_{1,n}\rangle }{f\langle g_{n}\rangle }I_{2}\gamma _{n}+\frac{iy}{f}%
I_{2}\gamma _{1,n}=-a_{1}\gamma _{n}+b_{1}\gamma _{1,n},  \label{4.3.1}
\end{equation}
where $I_{2,}\ a_{1}$ and $b_{1}$ are defined by formulas (\ref{3.12.1}) and
(\ref{3.12.2}), in which we have to replace $\Delta _{1}^{\prime },$ $\Delta
_{1}^{^{\prime \prime }},$ $f^{\prime }$ and $f^{\prime \prime }$ by $\Delta
_{1},$ $\langle \delta _{1,n}\rangle ,$ $f$ and $\langle g_{n}\rangle $
respectively. Denote by $\Phi =\{\Phi _{ij}\}_{i,j=1}^{3}$ the matrix
defined by the l.h.s. of system (\ref{ls}) and by $\Gamma =\{\Gamma
_{i}\}_{i=1}^{3}$ the vector with components $\Gamma _{1}=\gamma _{n},$ $%
\Gamma _{2}=\gamma _{1,n},$ $\Gamma _{3}=\gamma _{2,n}.$ Then we have from (%
\ref{Exp1}), (\ref{eps_r}) and (\ref{4.3.1})
\begin{equation}
\mathbf{E}\{|(\Phi \Gamma )_{1}|\}\leq \mathbf{E}\{|\varepsilon _{1,n}|\}.
\label{4.3.2}
\end{equation}
Interchanging in the above arguments indices $1$ and $2$ we obtain also that
\begin{equation}
\mathbf{E}\{|(\Phi \Gamma )_{2}|\}\leq \mathbf{E}\{|\varepsilon _{2,n}|\}.
\label{4.3.3}
\end{equation}
Besides, applying to the identity $G(z)(H_{1}+H_{2}-z)=1$ the operation $%
\langle n^{-1}Tr...\rangle $ and subtracting from the result the third
equation of system (\ref{2.4}), we obtain the one more relation

\begin{equation}
\mathbf{E}\{|(\Phi \Gamma )_{3}|\}=0.  \label{4.3.4}
\end{equation}
It follows from the proof of Proposition \ref{pr:3.3}
that the matrix $\Phi$ is
invertible if $y_{1}$ is big enough. Denote by $||...||_{1}$ the $l^{1}$%
-norm of $\mathbb{C}^{3}$ and by $||...||$ the induced matrix norm. Then we
have
\begin{equation}
\mathbf{E}\{||\Gamma ||_{1}\}\leq \mathbf{E}\{||\Phi ^{-1}\Phi \Gamma
||_{1}\}\leq \mathbf{E}^{1/2}\{||\Phi ^{-1}||^{2}\}\mathbf{E}^{1/2}\{||\Phi
\Gamma ||_{1}^{2}\}.  \label{4.3.5}
\end{equation}
It follows from our arguments above that all entries of the matrices $\Phi $
and $\Phi ^{-1}$ and all components of the vector $\Gamma $ are bounded
uniformly in $n$ and in realizations of random matrices $A_{n}$,$B_{n},U_{n}$
and $V_{n}$ in (\ref{2.1}). Thus we have
\[
||\Phi ^{-1}||\leq \sum_{i,j=1}^{3}|(\Phi ^{-1})_{ij}|\leq C,\;||\Phi \Gamma
||_{1}\leq \sum_{i,j=1}^{3}|\Phi _{ij}||\Gamma |_{j}\leq C.
\]

These bounds and (\ref{4.3.2}) - (\ref{4.3.5}) imply that
\[
\mathbf{E}\{||\Gamma ||_{1}\}\leq C^{3/2}(\mathbf{E}\{|\varepsilon _{2,n}|\}+%
\mathbf{E}\{|\varepsilon _{2,n}|\})^{1/2}.
\]
In view of (\ref{eps_r}) this inequality imply (\ref{Exp}), i.e. the
assertion of the lemma.$\blacksquare $

\bigskip

Now we extend the result of Lemma \ref{l:4.1}
for the case of unbounded $A_{n}$ and $%
B_{n}$, having the limiting NCM's with the finite first moments. We will
apply the standard in probability truncation technique, whose random matrix
version was used already in \cite{Ma-Pa:67,Pa:72}. \bigskip

\textit{Proof of Theorem \ref{th:main}}.
Without loss of generality we can assume that
\begin{equation}
\sup_{n}\int |\lambda |\mathbf{E}\{N_{1,n}(\mathrm{d}\lambda )\}\leq
m_{1}<\infty .  \label{4.23}
\end{equation}
For any $T>0$ introduce the matrices $A_{n}^{T}$ and $B_{n}^{T}$ replacing
eigenvalues $A_{n}$ and $B_{n}$ lying in $]T,\infty \lbrack $ by $T$ and
eigenvalues lying in $]-\infty ,-T]$ by $-T$. Denote by $N_{r,n}^{T},\ r=1,2$
the NCM of $A_{n}^{T}$ and $B_{n}^{T}$. It is clear that for any $T >0$ and
$r=1,2$, the sequence $\{N_{r,n}^{T}\}_{n\ge 1}$
converge weakly in probability to the measures $N_{r}^{T}$ as $n\rightarrow
\infty $, where $N_{r}^{T}$ are analogously defined via $N_{r}$ and have
their supports in $[-T,T]$, and that for each $r=1,2$ the sequence
$\{N_{r}^{T}\}_{T\ge 1}$ converge weakly to $N_{r}$
as $T\rightarrow \infty $. Denote by $N_{n}^{T},\ r=1,2$ the NCM of $%
H_{n}^{T}=H_{1,n}^{T}+H_{2,n}^{T}=V_{n}^{\ast }A_{n}^{T}V_{n}+U_{n}^{\ast
}B_{n}^{T}U_{n}$. According to linear algebra, if $M_{r},r=1,2$ are two
Hermitian $n\times n$ matrices, then
\begin{equation}
\mathrm{rank}(M_{1}+M_{2})\leq \mathrm{rank}M_{1}+\mathrm{rank}M_{2},
\label{rank}
\end{equation}
and if $\{\mu _{r,l}\}_{l=1}^{n},r=1,2$ are eigenvalues of $M_{r},r=1,2$,
then for any Borel set $\Delta \in \mathbb{R}$%
\[
|\#\{\mu _{1,l}\in \Delta \}-\#\{\mu _{2,l}\in \Delta \}|\leq \mathrm{rank}%
(M_{1}-M_{2}).
\]
By using these facts we find that
\begin{equation}\begin{array}{c}
|N_{n}(\Delta )-N_{n}^{T}(\Delta )| \leq \frac{1}{n}\mathrm{rank}%
(H_{n}-H_{n}^{T})\leq \frac{1}{n}\mathrm{rank}(A_{n}-A_{n}^{T})+\\
+\frac{1}{n}\mathrm{rank}(B_{n}-B_{n}^{T})\leq
N_{1,n}(\mathbb{R}\backslash ]-T,T[)+N_{2,n}(\mathbb{R}\backslash ]-T,T[),
\end{array}       \label{N^T}
\end{equation}
valid for any Borel set $\Delta \in \mathbb{R}$. As a result, the Stieltjes
transform $g_{n}^{T}$ of $N_{n}^{T}$ and the Stieltjes transform $g_{n}$ of $%
N_{n}$ are related as follows:
\[
|g_{n}^{T}(z)-g_{n}(z)|\leq \frac{\pi }{|\mathrm{Im}z|}\left( N_{1,n}(%
\mathbb{R}\backslash ]-T,T[)+N_{2,n}(\mathbb{R}\backslash ]-T,T[)\right) ,
\]
hence
\begin{equation}
\mathbf{E}\{|g_{n}^{T}(z)-g_{n}(z)|\}\leq \frac{\pi }{|\mathrm{Im}z|}\left(
\mathbf{E\{}N_{1,n}(\mathbb{R}\backslash ]-T,T[)\}+\mathbf{E\{}N_{2,n}(%
\mathbb{R}\backslash ]-T,T[)\}\right) .  \label{4.26}
\end{equation}
and
\[
\lim_{n\rightarrow \infty }\mathbf{E}\{N_{r,n}(\mathbb{R}\setminus
]-T,T[)\}\leq 1-N_{r}(]-T,T[)=o(1),\quad T\rightarrow \infty .
\]

Since the norms of matrices $H_{1}^{T}$ and $H_{2}^{T}$ are bounded, the
results of the Lemma \ref{l:4.1}
are applicable to the function $g_{n}^{T}(z)$, so
that, in particular, for any non-real $z$ it converges in probability as $%
n\rightarrow \infty $ to a function $f^{T}(z)$ satisfying the system
\[
\begin{array}{rl}
f^{T}(z) & =f_{1}^{T}\displaystyle\left( z-\frac{\Delta _{2}^{T}(z)}{f^{T}(z)%
}\right) \\
f^{T}(z) & =f_{2}^{T}\displaystyle\left( z-\frac{\Delta _{1}^{T}(z)}{f^{T}(z)%
}\right) \\
f^{T}(z) & =\displaystyle\frac{1-\Delta _{1}^{T}(z)-\Delta _{2}^{T}(z)}{-z}
\end{array}.
\]
In addition, since $\mathbf{E}\{g_{n}^{T}(z)\}$ and $\mathbf{E}\{\delta
_{1,n}^{T}(z)\}$ are bounded uniformly in $n$ and $T$ for $z\in E(y_{0}):$
\[ \begin{array}{l}
|\mathbf{E}\{g_{n}^{T}(z)\}|\leq {\frac{1}{y_{0}}},\\ |\mathbf{E}\{\delta
_{1,n}^{T}(z)\}|\leq {\frac{1}{y_{0}}}\int |\lambda |\mathbf{E}\{N_{1,n}^{T}(%
\mathrm{d}\lambda )\}\leq {\frac{1}{y_{0}}}\int |\lambda |\mathbf{E}%
\{N_{1,n}(\mathrm{d}\lambda )\}\leq {\frac{m_{1}}{y_{0}},}\end{array}
\]
we have
\begin{equation}
|f^{T}(z)|\leq {\frac{1}{y_{0}}},\ |\Delta _{1}^{T}(z)|\leq {\frac{m_{1}}{%
y_{0}}}  \label{4.27}
\end{equation}
Thus, there exists a sequence $T_{k}\rightarrow \infty $ such that sequences
of analytic functions $\{f^{T_{k}}(z)\}$ and $\{\Delta _{1}^{T_{k}}(z)\}$
converge uniformly on any compact of the $E(y_{0})$ of (\ref{4.26}). In
addition, the measures $N_{r}^{T_{k}},r=1,2$ converge weakly to the limiting
measures $N_{r},r=1,2$. Hence, there exist three analytic functions $f(z)$, $%
\Delta _{1}(z)$ and $\Delta _{2}(z)=zf(z)+1-\Delta _{1}(z)$ verifying (\ref
{2.4}). Besides, because of (\ref{4.27}) and (\ref{3.1}) for $z\in E(y_{0})$
we have
\[
|\Delta _{1}(z)|\leq {\frac{m_{1}}{y_{0}}},\mathrm{and\;}\Delta
_{2}(z)=o(1)\ \mathrm{as}\ y_{0}\rightarrow \infty .
\]
As a result of relations above, $f(z)$ and $\Delta _{r}(z),r=1,2$ satisfy
the conditions of Proposition \ref{pr:3.3}, hence they are defined uniquely.

Furthermore, we have
\[
\mathbf{E}\{|g_{n}(z)-f(z)|\}\leq \mathbf{E}\{|g_{n}(z)-g_{n}^{T_{k}}(z)|\}+%
\mathbf{E}\{|g_{n}^{T_{k}}(z)-f^{T_{k}}(z)|\}+|f^{T_{k}}(z)-f(z)|.
\]
Hence in view of (\ref{4.26}), arguments above on convergence of $f^{T_{k}}$
to $f$ , and Lemma \ref{l:4.1} we conclude that for each $z\in E(y_{0})$%
\[
\lim_{n\rightarrow \infty }\mathbf{E}\{|g_{n}(z)-f(z)|\}=0.
\]
In view of Proposition \ref{pr:4.1} this implies that the NCM (\ref{NCM})
of random
matrices (\ref{2.1}) converges weakly in probability as $n\rightarrow \infty
$ to the non-random measure, whose Stieltjes transform is a unique solution
of system (\ref{2.4}).$\blacksquare $

\setcounter{equation}{0}
\setcounter{theorem}{0}
\setcounter{proposition}{0}
\setcounter{lemma}{0}
\setcounter{remark}{0}
\section{Properties of the Solution}

Here we will consider several simple properties of the limiting eigenvalue
counting measure described by Theorem \ref{th:main},
i.e. the measure, whose Stieltjes
transform is a solution of (\ref{2.4}) satisfying (\ref{bound})--(\ref{as}).
We refer the reader to works \cite{Vo-Dy-Ni:92,Be-Vo:93,Bi:97,Be-Vo:98}
and references therein for a rather complete collection of results on
properties of the measure, resulting from the binary operation in the space
of the probability measures, defined by a version of system (\ref{2.4}).
This binary operation is called free additive convoluton.

\bigskip

(i) \textit{Assume that the supports of the limiting eigenvalue measures of
the matrices }$A_{n}$\textit{\ and }$B_{n}$\textit{\ are bounded, i.e. there
exist }$-\infty <a_{r,}\;b_{r}<\infty ,r=1,2,$\textit{\ such that}
\begin{equation}
\mathrm{supp}\,\,N_{r}\subset \lbrack a_{r},b_{r}],r=1,2.  \label{supNr}
\end{equation}
\textit{Then}
\begin{equation}
\mathrm{supp}\,N\subset \lbrack a_{1}+a_{2},b_{1}+b_{2}].  \label{supN}
\end{equation}

\medskip

\textit{Proof}. Denote by $\{\lambda _{l}\}_{l=1}^{n}$ and by $\{\lambda
_{r,l}\}_{l=1}^{n},r=1,2$ eigenvalues of $H_{n}$ and $H_{r,n}$ in (\ref{2.1}%
) respectively. Then, according to the linear algebra (cf.(\ref{N^T})),
\[
\#\{\lambda _{l}\in \mathbb{R}\backslash \lbrack
a_{1}+a_{2},b_{1}+b_{2}]\}\leq \#\{\lambda _{1,l}\in \mathbb{R}\backslash
\lbrack a_{1},b_{1}]\}+\#\{\lambda _{2,l}\in \mathbb{R}\backslash \lbrack
a_{2},b_{2}]\}.
\]
In view of Theorem \ref{th:main}
and (\ref{supNr}) this leads to the relation $N(%
\mathbb{R}\diagdown \sigma )=0$, i.e. to (\ref{supN}).

\bigskip

(ii). \textit{Examples.} 1. Consider the case when $A_{n}\mathit{=}B_{n}$,
i.e. $N_{1}=N_{2}$. In this case system (\ref{2.4}) will have the form
\begin{equation}
f(z)=f_{1}\left( \frac{z}{2}-\frac{2}{f(z)}\right) .  \label{5.3}
\end{equation}

Take $N_{1}=N=\alpha $ $\delta _{0}+(1-\alpha )$ $\delta _{a}$ where $0\leq
\alpha \leq 1$, $a>0$ and $\delta _{\lambda }$ is the unit measure
concentrated at $\lambda \in \mathbb{R}$. Then
\[
f_{1}(z)=\frac{-\alpha }{z}+\frac{1-\alpha }{a-z}
\]
and (\ref{2.4}) reduces to the quadratic equation
\[
z(z-2a)f^{2}+2a(1-2\alpha )f-1=0,
\]
whose solution satisfying (\ref{bound}) - (\ref{as}) is
\[
f(z)=\frac{-a(1-2\alpha )-\sqrt{(z-\lambda _{+})(z-\lambda _{-})}}{z(z-2a)}%
,\;\lambda _{\pm }=a(1\pm 2\sqrt{\alpha (1-\alpha )}).
\]
By using (\ref{FP}) we find that the limiting measure in this case has the
form
\begin{equation}
N=(2\alpha -1)_{+}\delta _{0}+(1-2\alpha )_{+}\delta _{2a}+N^{\ast },
\label{5.4}
\end{equation}
where $x_{+}=\max (0,x),$ and
\begin{equation}
N^{\ast }(\mathrm{d}\lambda )=\frac{1}{\pi }\frac{\sqrt{(\lambda
_{+}-\lambda )(\lambda -\lambda _{-})}}{\lambda (\lambda -2a)}\chi _{\lbrack
0,2a]}(\lambda )\mathrm{d}\lambda  \label{5.5}
\end{equation}
is the absolute continuous measure of the mass $1-2\alpha .$ Here $\chi
_{\Delta }(\lambda )$ is the indicator of the set $\Delta \subset \mathbb{R}$%
. In the cases $\alpha =0,1$ (\ref{5.4}) is $\delta _{2a}$ and $\delta _{0}$
respectively, and in the case $\alpha =1/2$ (\ref{5.4}) has no atoms, but
only the square root singularities
\begin{equation}
N^{\ast }(\mathrm{d}\lambda )=\frac{1}{\pi \sqrt{\lambda (2a-\lambda )}}\chi
_{\lbrack 0,2a]}(\lambda )\mathrm{d}\lambda  \label{5.6}
\end{equation}
Formulas (\ref{5.3})--(\ref{5.6}) shows that:

\begin{itemize}
\item  the result (\ref{supN}) is optimal with respect to the endpoints of
the measures $N_{r},r=1,2$ and $N$;

\item  in the case when $N_{1}=N_{2}$ have atoms of the mass $\mu >1/2$ at
the same point then the measure $N$ has also an atom of the mass $(2\mu -1)$
(for general results of this type see \cite{Be-Vo:98}).
\end{itemize}

\bigskip

However, in general the support of $N$ is strictly included in the sum of
supports of measures $N_{r},r=1,2$, i.e. the inclusion in the r.h.s part of (%
\ref{5.3}) is strict. This can be illustrated by the following two examples.

\bigskip
2. Take again $N_{1}=N_{2}$, where now
\[
N_{1}(\mathrm{d}\lambda )=\frac{1}{\pi \sqrt{a^{2}-\lambda ^{2})}}\chi
_{\lbrack -a,a]}(\lambda )\mathrm{d}\lambda .
\]
is the arcsin law. This measure corresponds to the matrix ensemble (\ref{MM}%
) with
\begin{equation}
V(\lambda )=\left\{
\begin{array}{lll}
0, & |\lambda |<1, &  \\
\infty , & |\lambda |>1. &
\end{array}
\right.  \label{Leg}
\end{equation}
In this case equation (\ref{5.3}) is again quadratic and leads to
\[
N(\mathrm{d}\lambda )=\frac{\sqrt{3a^{2}-\lambda ^{2})}}{\pi (4a^{2}-\lambda
^{2})}\chi _{\lbrack -\sqrt{3}a,\sqrt{3}a]}(\lambda )\mathrm{d}\lambda .
\]
\bigskip

3. In the next example we take
\[
N_{r}(\mathrm{d}\lambda )=\frac{1}{4\pi a_{r}^{2}}\sqrt{8a_{r}^{2}-\lambda
^{2}}\chi _{\lbrack -2\sqrt{2}a_{r},2,\sqrt{2}a_{r}]}(\lambda )\mathrm{d}%
\lambda ,r=1,2,
\]
i.e. the both measures are the semicircle laws (\ref{Wi}). Then it is easy
to find that $N$ is also the semicircle measure with the parameter $a^{2}=$ $%
a_{1}^{2}+a_{2}^{2}.$ This case was indicated in \cite{Pa:72}. It can be
easily deduced from the law of addition of the R-transforms of Voiculescu
\cite{Vo-Dy-Ni:92}, because in this case $R_{r}(f)=2a_{r}^{2}f$. For further
properties of the measure $N$ in the case when one of $N_{r},r=1,2$ is the
semicircle law see \cite{Pa-Kh:93,Bi:97}.

\bigskip
(iii). \textit{Suppose that one of the measures $N_{r}(\mathrm{d}\lambda
),r=1,2$ is absolute continuous with respect to the Lebesgue measure, i.e.,
say, $N_{1}(\mathrm{d}\lambda )=\rho _{1}(\lambda )\mathrm{d}\lambda ,$ and}
\[
\overline{\rho }_{1}\mathit{=\mathrm{ess}\sup_{\lambda \in \mathbb{R}}|\rho
_{1}(\lambda )|<\infty ,}.
\]
\textit{\ Then $N$ is also absolute continuous with respect to the Lebesgue
measure, i.e.}$N(\mathrm{d}\lambda )=\rho (\lambda )\mathrm{d}\lambda ,$
\textit{and }
\begin{equation}
\mathrm{ess}\sup_{\lambda \in \mathbb{R}}|\rho _{1}(\lambda )|=\overline{%
\rho }_{1}<\infty .  \label{rho}
\end{equation}

\textit{Proof}. Indeed, since the function $z_{1}^{\ast }=z-\Delta
_{2,1}/f(z)$ is analytic for non-real $z$, the number of its zeros in any
compact of $\mathbb{C}\backslash\mathbb{R}$ is finite.
Thus, for any $\lambda \in\mathbb{R}$
there exists a sequence $\{z_{n}\}$ of non-real numbers such
that $z_{n}\rightarrow \lambda $ as $n\rightarrow \infty $ and $\mathrm{Im}$
$z_{n}^{\ast }\neq 0.$ Hence, we have from the first equation of system (\ref
{2.4}) for $z_{n}^{\ast }=$ $\lambda _{n}^{\ast }+i\varepsilon _{n}^{\ast }$
\[
\frac{1}{\pi }\mathrm{Im}f(z)=\frac{1}{\pi }\int \frac{\varepsilon
_{r}^{\ast }\rho _{r}(\mu )\mathrm{d}\mu }{(\mu -\lambda _{r}^{\ast
})^{2}+(\varepsilon _{r}^{\ast })^{2}}\leq \overline{\rho }_{1}\frac{1}{\pi }%
\int \frac{\varepsilon _{r}^{\ast }\mathrm{d}\mu }{(\mu -\lambda _{r}^{\ast
})^{2}+(\varepsilon _{r}^{\ast })^{2}}=\overline{\rho }_{1}.
\]
This relation and the inversion formula (\ref{FP}) yield (\ref{rho}). For
more general results in this direction see the recent paper \cite{Be-Vo:98}.

\setcounter{equation}{0}
\setcounter{theorem}{0}
\setcounter{proposition}{0}
\setcounter{lemma}{0}
\setcounter{remark}{0}
\section{Discussion}

In this Section we comment on several topics related to those studied above.
\medskip

1. In this paper we deal with Hermitian and unitary matrices, i.e. we assume
that the matrices $A_{n}$ and $B_{n}$ in (\ref{2.1}) are Hermitian and $%
U_{n} $ and $V_{n}$ are unitary. It is natural also to consider the case of
real symmetric $A_{n}$ and $B_{n}$ and orthogonal $U_{n}$ and $V_{n}$. This
case can be handled by using the analogue of formula (\ref{3.12}) of the
orthogonal group $O(n)$. Indeed, it is easy to see that \ this analogue has
the form
\[
\int\limits_{\mathit{O}(n)}\Phi ^{^{\prime }}(O^{T}MO)\cdot \lbrack
X,O^{T}MO]\mathrm{d}O=0,
\]
where $O^{T}$\ is the transposed to $O$ and $X$ is a real symmetric matrix.
By using this formula we obtain instead of (\ref{3.16.1})
\[
\langle G_{aa}(H_{2}G)_{bc}\rangle +\langle G_{ab}(H_{2}G)_{ac}\rangle
=\langle (GH_{2})_{aa}G_{bc}\rangle +\langle (GH_{2})_{ab}G_{bc}\rangle .
\]
The second terms in both sides of this formula give two additional terms
\[
-n^{-1}G^{T}H_{2}G+n^{-1}H_{2}G^{T}G.
\]
in the analogue of (\ref{3.241}). These terms, however, produce the
asymptotically vanishing contribution because, in view of (\ref{3.4}), (\ref
{3.7}) and (\ref{3.21.3}), we have
\[
\left| n^{-2}\langle \mathrm{Tr}(1+\Delta
_{2,n}f_{n}^{-1}G_{1})^{-1}G_{1}(-G^{T}H_{2}G+H_{2}G^{T}G)\rangle \right|
\leq \frac{2}{ny_{0}^{3}}m_{4}^{1/4}.
\]
Similar and also negligible as $n\rightarrow \infty $ terms appear in
analogues of formulas (\ref{3.35}), (\ref{3.47.1}) and (\ref{3.49}) of the
proof of Theorem \ref{th:3.2}.
As the result, we obtain in this case the same system (%
\ref{2.4}), defining the Stieltjes transform of the limiting eigenvalue
counting measure of the analogue of (\ref{2.1}) with the real symmetric $%
A_{n}$ and $B_{n}$ and orthogonal Haar-distributed $U_{n}$ and $V_{n}$.

\medskip

2. As was mentioned in the Introduction, our main result, Theorem
\ref{th:main}, can
be viewed as an extension of the result by Speicher \cite{Sp:93}, obtained
by the moment method and valid for uniformly in $n$ bounded matrices $A_{n}$
and $B_{n}$ in (\ref{2.1}). Both results are analogues for randomly rotated
matrices of old results of \cite{Ma-Pa:67,Pa:72} (see (\ref{fMP}) and (\ref
{dsc})) on the form of the limiting eigenvalue counting measure of the sum
of an arbitrary matrix and certain random matrices (see (\ref{MP}) and (\ref
{dGUE})), in particular, Gaussian random matrices (\ref{GUE}). In this case,
however, there exists another model, proposed by Wegner \cite{We:79} that
combines properties of random matrices, having all entries roughly of the
same order, and of random operators, whose entries decay sufficiently fast
in the distance from the principal diagonal (see e.g. \cite{Pa-Fi:92}). A
simple, but rather non-trivial version of the Wegner model corresponds to
the selfadjoint operator $H$ acting in $l^{2}(\mathbb{Z}^{d})\times \mathbf{C%
}^{n}$ and defined by the matrix
\begin{equation}
H(x,j;y,k)=v(x-y)\delta _{jk}+\delta (x-y)f_{jk}(x)  \label{We}
\end{equation}
where $x,y\in \mathbb{Z}^{d}$, $j,k=1,...,n$, $\delta (x)$ is the $d$%
-dimensional Kronecker symbol,
\begin{equation}
v(-x)=\overline{v}(x),\ \ \ \sum_{x\in \mathbb{Z}^{d}}|v(x)|<\infty ,
\label{v}
\end{equation}
and $f(x)=\{f_{jk}(x)\}_{j,k=1}^{n}$, $x\in \mathbb{Z}^{d}$ are independent
for different $x$ and identically distributed $n\times n$ Hermitian
matrices, whose distribution for any $x$ is given by (\ref{GUE}). According
to \cite{We:79} (see also \cite{Pa-Kh:93}) asymptotic for $n\rightarrow
\infty $ properties of operator (\ref{We}) resemble, in many aspects,
asymptotic properties of matrices (\ref{GUE}). The ''free'' analogue of the
Wegner model was proposed in \cite{Ne-Sp:95}. In this case i.i.d. matrices $%
f(x)$ have the form
\begin{equation}
f(x)=U_{n}^{\ast }(x)B_{n}U_{n}(x)  \label{fU}
\end{equation}
where $B_{n}$ is as in (\ref{2.1}) and $U_{n}(x)$, $x\in \mathbb{Z}^{d}$ are
i.i.d. unitary $n\times n$ matrices whose distribution is given by the Haar
measure on $\mathit{U}(n)$. By using a version of the moment method, similar
to that of paper \cite{Sp:93}, or, rather, its formal scheme, the authors
derived the limiting form of
\[
\mathbf{E}\left\{ n^{-1}\sum_{j=1}^{n}G(x,j;y,j)\right\} ,
\]
where $G(x,j;y,k)$ is the matrix (the Green function) of the resolvent $%
(H-z)^{-1}$ of (\ref{We}) - (\ref{fU}). The authors also found a certain
second moment of the Green function. This moment is necessary to compute the
a.c. conductivity via the Kubo formula. Because of the moment method results
of \cite{Ne-Sp:95} are valid for uniformly bounded in $n$ matrices $B_{n}$
in (\ref{fU}), similar to results for matrices (\ref{2.1}) obtained in \cite
{Sp:93}. By using a natural extension of the differentiation formula (\ref
{3.12}) and the technique developed in \cite{Pa-Kh:93} to analyze the Wegner
model, the results of paper \cite{Ne-Sp:95} can be extended to the case of
arbitrary matrices $B_{n}$ in (\ref{fU}), because in this case the role of
condition (\ref{2.2}) of Theorem \ref{th:main}) plays condition (\ref{v}).

\medskip

3. As was mentioned before asymptotic properties of random matrices are of
considerable interest in the certain branches of the operator algebra theory
and related branch of the non-commutative probability theory, known as free
probability (see \cite{Vo:91,Vo-Dy-Ni:92,Vo:98} and references therein).
Here large random matrices is an important example of the asymptotically
free non-commutative random variables, providing a sufficiently reach
analytic model of the abstract notion of freeness of elements of an operator
algebra. The most widely used examples of asymptotically free families of
non-commutative random variables are Gaussian random matrices and unitary
Haar-distributed random matrices. The proof of asymptotic freeness of
unitary matrices given in \cite{Vo:91,Vo-Dy-Ni:92} reduces to that for
complex Gaussian matrices basing on the observation that the unitary part of
the polar decomposition of complex Gaussian matrix with independent entries
is the Haar-distributed unitary matrix. This method requires certain
technicalities because of the singularity of the polar decomposition at
zero. On the other hand, the differentiation formula (\ref{3.12}) allows one
to prove directly similar statements. Here is an example of results of this
type (related results are proved in \cite{Xu:97}).

\begin{theorem}
\label{th:6.1} Let $k$ be a
positive integer, $\{T_{r,n}\}_{r=1}^{k}$ be a set of $n\times n$%
 matrices, such that
\begin{equation}
\sup_{r\leq k;\ k,l,n\in \mathbb{N}}n^{-1}\mathrm{Tr}(T_{r,n}^{\ast
}T_{r,n})^{l}<\infty ,  \label{supT}
\end{equation}
and let $U_{n}$ be the unitary and Haar-distributed random
matrix. If for any $k\in \mathbb{N}$%
\begin{equation}
\lim_{n\rightarrow \infty }n^{-1}\mathrm{Tr}T_{r,n}=0,\ \ r=1,...,k,
\label{T0}
\end{equation}
then for any set of non-zero integers such that $%
\{m_{r}\}_{r=1}^{k} $, $\sum_{r=1}^{k}m_{r}=0$ 
\begin{equation}
\lim_{n\rightarrow \infty }\langle n^{-1}\mathrm{Tr}%
U_{n}^{m_{1}}T_{1,n}...U_{n}^{m_{k}}T_{k,n}\rangle =0,  \label{res}
\end{equation}
where $\langle \cdot \rangle $ denotes the integration
with respect to the Haar measure over $U(n)$.
\end{theorem}

\begin{remark}{\rm The theorem is trivially true in the case when $%
\sum_{r=1}^{k}m_{r}\neq 0$.
}\end{remark}

In the two subsequent lemmas we omit the subindex $n$.

\begin{lemma}
\label{l:6.1} Let $\{T_{i}\}_{i=1}^{k}$ be a set of $n\times n$
matrices and $U$ is the Haar-distributed unitary
matrix. Then for any set of non-zero integers $\{m_{i}\}_{i=1}^{k}$, $%
\sum_{i=1}^{k}m_{i}=0$ the following identity holds:
\[
{n^{-1}}\mathrm{Tr}\langle U^{m_{1}}T_{1}...U^{m_{k}}T_{k}\rangle
=-\sum_{l_{1}=2}^{m_{1}}\langle {n^{-1}}\mathrm{Tr}U^{l_{1}-1}{n^{-1}}%
\mathrm{Tr}(U^{m_{1}-l_{1}+1}T_{1}...U^{m_{k}}T_{k})\rangle -
\]
\begin{equation}
\sum_{r\in \{2,...,k\},m_{r}>0\ }\sum_{l_{r}=1}^{m_{r}}\langle {n^{-1}}%
\mathrm{Tr}(U^{m_{1}}T_{1}...T_{r-1}U^{l_{r}-1}){n^{-1}}\mathrm{Tr}%
(U^{m_{r}-l_{r}+1}T_{r}...U^{m_{k}}T_{k})\rangle +  \label{6.1}
\end{equation}
\[
\sum_{r\in \{2,...,k\},m_{r}<0\ }\sum_{l_{r}=1}^{-m_{r}}\langle {n^{-1}}%
\mathrm{Tr}(U^{m_{1}}T_{1}...T_{r-1}U^{-l_{r}}){n^{-1}}\mathrm{Tr}%
(U^{m_{r}+l_{r}}T_{r}...U^{m_{k}}T_{k})\rangle .
\]
\end{lemma}
\smallskip

\textit{Proof.} Without loss of generality assume that $m_{1}>0$. Then,
using the analogue of formula (\ref{3.12}) for the average $\langle \lbrack
U^{m_{1}}T_{1}...U^{m_{k}}T_{k}]_{ab}\rangle $, we obtain for any Hermitian $%
X$%
\[
\sum_{r\in \{1,...,k\},m_{r}>0}\sum_{l_{r}=1}^{m_{r}}\langle \lbrack
U^{m_{1}}T_{1}...T_{r-1}U^{l_{1}-1}XU^{m_{r}-l_{r}+1}T_{r}...U^{m_{k}}T_{k}]_{ab}\rangle +
\]
\begin{equation}
\sum_{r\in \{2,...,k\},m_{r}<0}\sum_{l_{r}=1}^{-m_{r}}\langle \lbrack
U^{m_{1}}T_{1}...T_{r-1}U^{-l_{r}}XU^{m_{r}+l_{r}}T_{r}...U^{m_{k}}T_{k}]_{ab}\rangle =0
\label{6.2}
\end{equation}
Choosing as $X$ the Hermitian matrix having only $(c,d)$-th and $(d,c)$-th
non-zero entries, setting then $a=c$ and $\ b=d$ and applying to the result
the operation $n^{-2}\sum_{a,b}$, we obtain (\ref{6.1}).$\blacksquare $

\begin{lemma}
\label{l:6.2} Under the conditions (\ref{supT}) and (%
\ref{T0}) the variance $D=\langle |\xi ^{\circ }|^{2}\rangle $ of
the random variable
\begin{equation}
\mathit{\xi ={n^{-1}}\mathrm{Tr\mathit{L,\;L}=}%
U^{m_{1}}T_{1}...U^{m_{k}}T_{k}}  \label{xiL}
\end{equation}
is of the order $n^{-2}$ as $n\rightarrow \infty$.
\end{lemma}
\medskip

\textit{Proof.} Using the same technique as that in Lemma \ref{l:6.1}
for $\langle
\overline{L_{ab}}L_{cd}\rangle $ we obtain the relation
\[
D=-\sum_{l_{1}=2}^{m_{1}}\langle \overline{{\xi }}^{\circ }{n^{-1}}\mathrm{Tr%
}U^{l_{1}-1}{n^{-1}}\mathrm{Tr}(U^{m_{1}-l_{1}+1}T_{1}...U^{m_{k}}T_{k})%
\rangle -
\]
\begin{equation}
\sum_{r\in \{2,...,k\}, m_{r}>0}
\sum_{l_{r}=1}^{m_{r}}\langle \overline{{\xi }}^{\circ }{n^{-1}}%
\mathrm{Tr}(U^{m_{1}}T_{1}...T_{r-1}U^{l_{r}-1}){n^{-1}}\mathrm{Tr}%
(U^{m_{r}-l_{r}+1}T_{r}...U^{m_{k}}T_{k})\rangle +  \label{6.3}
\end{equation}
\[
\begin{array}{c}
\sum\limits_{r\in \{2,...,k\},
m_{r}<0}\sum\limits_{l_{r}=1}^{-m_{r}}\langle
\overline{{\xi }}^{\circ }{n^{-1}}%
\mathrm{Tr}(U^{m_{1}}T_{1}...T_{r-1}U^{-l_{r}}){n^{-1}}\mathrm{Tr}%
(U^{m_{r}+l_{r}}T_{r}...U^{m_{k}}T_{k})\rangle\\ +{n^{-2}}\Phi ,
\end{array}
\]
where
\[
\Phi =-\sum_{r\in \{1,...,k\},
m_{r}>0\ }\sum_{l_{r}=1}^{m_{r}}{n^{-1}}{\mathrm{Tr}\langle
(U^{m_{r}-l_{r}+1}T_{r}...T_{k}U^{m_{1}}T_{1}...T_{r-1}U^{l_{r}-1}})^*
L\rangle +
\]
\[
+\sum_{r\in \{2,...,k\},
m_{r}<0\ }\sum_{l_{r}=1}^{-m_{r}}{n^{-1}} \mathrm{Tr}\langle
(U^{m_{r}+l_{r}}T_{r}...T_{k}U^{m_{1}}T_{1}...T_{r-1}U^{-l_{r}})^* L\rangle.
\]
We have obviously for $k=m=1$%
\[
\langle {n^{-1}}\overline{\mathrm{Tr}(UT)^{\circ }}{n^{-1}}\mathrm{Tr}%
(UT)\rangle \leq {\frac{1}{n^{2}}}{n^{-1}}\mathrm{Tr}(TT^{\ast }).
\]
We proceed further by induction. In view of condition (\ref{supT}) and
Proposition \ref{pr:3.1} we have the bound
\begin{equation}
|{n^{-1}}\mathrm{Tr}(U^{m_{1}}T_{r_{1}}...U^{m_{p}}T_{r_{p}})|\leq C_{,}^{2}
\label{6.3.1}
\end{equation}
where $C$ may depend only on $p$. Now, since ${n^{-1}}\mathrm{Tr}\langle
U^{l}\rangle =0,l\neq 0,$ the summands of the first term in r.h.s. of (\ref
{6.3}) can be estimated as follows
\begin{equation}
\left| \langle \overline{{\xi }}^{\circ }{n^{-1}}\mathrm{Tr}(U^{l_{1}}){%
n^{-1}}\mathrm{Tr}(U^{m_{1}-l_{1}+1}T_{1}...U^{m_{k}}T_{k})\rangle \right|
\leq C\sqrt{D}\sqrt{\langle |{n^{-1}}\mathrm{Tr}(U^{l_{1}})^{\circ
}|^{2}\rangle }.  \label{6.5}
\end{equation}
Likewise, by using the cyclic property of the trace, the identity $\langle
a^{\circ }bc\rangle =\langle a^{\circ }b^{\circ }c\rangle +\langle a^{\circ
}c^{\circ }\rangle \langle b\rangle $, Schwarz inequality, and (\ref{6.3.1}%
), we obtain for the second term in the right-hand side of (\ref{6.3}) the
following estimates for $r\geq 2$%
\begin{equation}\begin{array}{c}
\left| \langle \overline{{\xi }}^{\circ }{n^{-1}}\mathrm{Tr}%
(U^{m_{1}}T_{1}...T_{r-1}U^{l_{r}-1}){n^{-1}}\mathrm{Tr}%
(U^{m_{r}-l_{r}+1}T_{r}...U^{m_{k}}T_{k})\rangle \right| \leq\\
\leq C\sqrt{D}\left\{ \sqrt{\langle |{n^{-1}}\mathrm{Tr}%
(U^{m_{1}+l_{r}-1}T_{1}...U^{m_{r-1}}T_{r-1})^{\circ }|^{2}\rangle }+\right.\\
\left. +\sqrt{\langle |{n^{-1}}\mathrm{Tr}(U^{m_{r}-l_{r}+1}T_{r}...
U^{m_{k}}T_{k})^{\circ}|^{2}\rangle }\right\}\end{array}  \label{6.4}
\end{equation}
The third term in the right-hand side of (\ref{6.3}) can be estimated
analogously. The forth term is of the order $1/n^{2}$ in view of (\ref{xiL}%
). By the induction hypothesis the expectations under square roots in the
r.h.s. of (\ref{6.4}) and (\ref{6.5}) are of the order $n^{-2}$. This leads
to the inequality
\[
D\leq \frac{C_{1}}{n}\sqrt{D}+\frac{C_{2}}{n^{2}},
\]
where $C_{1}$ and $C_{2}$ are independent of $n$. This implies the bound $%
D=O(n^{-2})$.$\blacksquare $

\bigskip

\textit{Proof of Theorem \ref{th:6.1}}.
We use Lemma \ref{l:6.1} and again the induction. We
have first
\[
{n^{-1}}\mathrm{Tr}\langle U^{m}T_{1}U^{-m}T_{2}\rangle ={n^{-1}}\mathrm{Tr}%
T_{1}{n^{-1}}\mathrm{Tr}T_{2}=0.
\]
In general case we use Lemma \ref{l:6.2}
to factorize asymptotically the moments in
the r.h.s. of (\ref{6.1}). In the resulting relation the expressions ${n^{-1}%
}\mathrm{Tr}\langle U^{m_{r_{1}}}T_{r_{1}}...U^{m_{r_{s}}}T_{r_{s}}\rangle $
are zero for any collection $(T_{r_{1}},...,T_{r_{s}})$ and any $n$, if $%
\sum_{i=1}^{s}m_{r_{i}}\neq 0,$ and tend to zero as $n\rightarrow \infty $
if $\sum_{i=1}^{s}m_{r_{i}}=0$ in view of the induction hypothesis and
condition (\ref{T0}). This leads to (\ref{res}).$\blacksquare $

\begin{remark}
{\rm A simple version of the above arguments allows us to prove
that the normalized counting measure of the Haar distributed unitary
matrices converges with probability one to the uniform distribution on the
unit circle. Indeed, consider again the Stieltjes transform $g_{n}$ of this
measure, supported now on the unit circle. By the spectral theorem for
unitary matrices we have
\begin{equation}
g_{n}(z)=n^{-1}\mathrm{Tr}G(z),\ G(z)=(U-z)^{-1},\ |z|\neq 1.  \label{stun}
\end{equation}
We can then obtain the following identities
\begin{equation}
\langle \mathrm{Tr}G^{2}(z)U\rangle =0,\;\langle g_{n}(z)n^{-1}\mathrm{Tr}%
G(u)U\rangle =0,  \label{G2U}
\end{equation}
\begin{equation}
\langle g_{n}(z_{1})n^{-1}\mathrm{Tr}G(z_{1})Ug(z_{2})\rangle +\langle n^{-3}%
\mathrm{Tr}G(z_{1})G(z_{2})UG(z_{2})\rangle =0.  \label{cor}
\end{equation}
By using the obvious relations
\[
G^{\prime }(z)=G^{2}(z),\ G(0)=U^{-1},\ G(\infty )=0,
\]
we obtain from the first of identities (\ref{G2U})
\[
f_{n}(z)\equiv \langle g_{n}(z)\rangle =\left\{
\begin{array}{lll}
0, & |z|<1, &  \\
-z^{-1}, & |z|>1. &
\end{array}
\right. .
\]
This relation shows that the expectation of the normalized counting measure
of $U$ is the uniform distribution on the unit circle, the fact that follows
easily from the shift invariance of the Haar measure. Now the second
identity (\ref{G2U}) and (\ref{cor}) lead to the bound
\[
|\langle g_{n}(z)\rangle |^{2}\leq \frac{C(r_{0})}{n^{2}},\ |z|\leq r_{0,}
\]
where $C(r_{0})$ is independent of $n$ and finite if $r_{0}$ is small
enough. This bound and arguments analogous to those used in the proof of
Theorem \ref{th:3.1}
imply that the normalized eigenvalue counting measure of unitary
Haar distributed random matrices converges with probability one to the
uniform distribution on the unit circle. This fact as well as the
analogous fact for the orthogonal group can be deduced from the works
by Dyson (see e.g. \cite{Me:91}), where the joint probability distribution of
all $n$ eigenvalues of the Haar distributed unitary or orthogonal matrices
was found and studied. This technique is more powerful but also more complex
than that used above and based on rather elementary means.
}\end{remark}
\vskip 1cm

{\bf Acknowledgements.}
V. Vasilchuk is thankful to
Laboratoire de Physique Math\'{e}matique et G\'{e}om\'{e}trie de
l'Universit\'{e} Paris-7 for hospitality  and
to Minist{\`{e}}re des Affaires Etrang{\`{e}}res de France
for the financial support.

\end{document}